\documentclass[a4paper,11pt]{article}
\usepackage{jcappub}
\usepackage{lineno}
\usepackage{graphicx}
\usepackage{tikz-feynman}
\usepackage{slashed}
\usepackage{array}
\usepackage{multirow}
\usepackage{orcidlink}
\nolinenumbers

\DeclareUnicodeCharacter{2212}{-}

\title{Lepton Collider as a window to reheating via freezing out Dark Matter detection}

\author{Subhaditya Bhattacharya\,\orcidlink{0000-0002-8841-603X}\,}
\author{\!\!, Anupam Ghosh\,\orcidlink{0000-0003-4163-4491}\,}
\author{\!\!, Niloy Mondal\,\orcidlink{0009-0006-5837-9772}\,, and}
\author{Abhik Sarkar\,\orcidlink{0000-0003-1449-2934}\,}
\affiliation{Department of Physics, Indian Institute of Technology Guwahati,\\ North Guwahati, Assam-781039, India}

\emailAdd{subhab@iitg.ac.in, anupamg@rnd.iitg.ac.in, niloy18@iitg.ac.in, sarkar.abhik@iitg.ac.in}

\abstract{We investigate the possibility of probing the reheating era via detection of freezing out dark matter (DM) at the future lepton collider. We assume that DM interaction with the Standard Model (SM) occurs via leptophilic effective operator, having suppressed interaction with the quarks and gluons due to the absence of DM signal at the LHC and direct searches. In order to have the seed of reheating in the DM relic density, unlike the conventional WIMP scenarios, we consider DM freeze-out to occur during the prolonged reheating epoch, driven by the inflaton decay. This in turn alters the thermal evolution of the DM abundance, making it sensitive to the entropy injection. We focus on mono-Higgs plus missing energy signal at the future lepton colliders via a detailed signal-background analysis using both polarized and unpolarized beams. Our result illustrates how collider experiments, when interpreted jointly with cosmological constraints, can provide indirect hints towards the Universe's thermal history, offering potential insights to the reheating temperature and the dynamics preceding the Big Bang Nucleosynthesis (BBN).}

\begin{document}
\maketitle
\flushbottom

\section{Introduction}
\label{sec:intro}
The thermal history of the Universe prior to Big Bang Nucleosynthesis (BBN) remains largely unconstrained, allowing for non-standard cosmological phases such as reheating~\cite{Bassett:2005xm,Allahverdi:2010xz}, during which inflaton decay products dominate the energy density before the onset of radiation domination. In the reheating phase, the altered expansion rate and continuous entropy injection can significantly modify the thermal freeze-out of weakly interacting massive particles (WIMPs)~\cite{Roszkowski:2017nbc} dwelling in the early Universe. Unlike the standard scenario, WIMPs may decouple earlier due to the faster Hubble expansion, leading to an overabundance that is diluted later by entropy injection. As a result, the final relic density becomes sensitive to both the WIMP annihilation cross section and the reheating dynamics, particularly to the reheating temperature ($T_{\rm RH}$). This work examines a phenomenologically viable freeze-out scenario occurring during the reheating phase, highlighting how cosmological, collider, and other experimental constraints can jointly probe such non-standard WIMP production mechanism and provide insights towards the thermal history of the Universe.

Since inception, the Large Hadron Collider (LHC) has pursued a wide range of DM search strategies, primarily targeting WIMPs~\cite{CMS:2020ulv,CMS:2017jrd,CMS:2017jdm,CMS:2016gsl,CMS:2018ysw,ATLAS:2017bfj,ATLAS:2019wdu,ATLAS:2017hoo,ATLAS:2017nga}. On the theoretical front, extensive phenomenological studies have investigated the sensitivity of LHC across a broad class of DM models, and signals~\cite{Djouadi:2011aa,Kahlhoefer:2017dnp,Ghosh:2021noq,Ghosh:2024boo,Boveia:2016mrp,Ghosh:2023xhs,Ghosh:2022rta,Abdallah:2015ter,Belanger:2018sti,Lee:2018pag,Ghosh:2024nkj,Roy:2024yoh,Ghosh:2025agw,Blasi:2023hvb}. These searches typically exploit missing transverse energy ($\slashed{E}_T$) in association with visible objects like jets, fat jets, photons, or electroweak bosons and have placed stringent constraints on various theoretical models, often comparable to or even exceeding those from non-collider approaches such as direct and indirect detection experiments. Despite this extensive effort, no significant deviation from the Standard Model (SM) predictions has been observed. This lack of evidence suggests that DM may not couple to the QCD sector of the SM, shifting the attention toward scenarios where DM interacts predominantly with the lepton sector. Such connections can be realized either by introducing explicit portals between DM and SM leptons in specific New Physics (NP) models, or through higher-dimensional effective (EFT) operators in a model-independent way. We focus in particular to the leptophilic EFT framework. Also, it is easier to probe such operators in the proposed electron-positron collider, than at the future sensitivities of LHC, for variety of reasons, as we elaborate.


The leptophilic operator that we study faces stringent constraints within the standard WIMP paradigm. For DM masses below 70\,GeV, the parameter space yielding correct or under-abundant relic density is entirely excluded by existing experimental data. In contrast, non-standard freeze-out scenario still allow viable regions consistent with current bounds and capable of producing the correct relic abundance thanks to the dilution produced by the entropy injection.

We investigate the collider phenomenology of this scenario at the International Linear Collider (ILC) operating at center-of-mass (CM) energy of $\sqrt{s} = 1$ TeV with an integrated luminosity of 8 ab$^{-1}$. In particular, we focus on the mono-$h$ channel, where scalar DM particles are produced in association with a Higgs boson, leading to distinctive mono-Higgs plus missing energy signatures. We perform a detailed signal-background analysis using both polarized and unpolarized beams to assess the sensitivity of the ILC to the parameter space consistent with reheating-era freeze-out. This study thus demonstrates how collider searches could provide indirect insight into the reheating temperature, linking terrestrial experiments to the dynamics of the post-inflationary epoch. Similar studies have examined the interpretation of reheating temperature from collider signatures in the context of freeze-in (FIMP) DM~\cite{Barman:2024nhr,Barman:2024tjt,Borah:2025ema}, but this is the first example of providing a similar insight via freezing out DM or WIMP.

The paper is organized as follows: In Sec.~\ref{sec:dmprod}, we discuss the dynamics of the early Universe and their impact on DM production. In Sec.~\ref{sec:constraints}, we review existing constraints on leptophilic DM from observational and experimental data. In Sec.~\ref{sec:collider}, we demonstrate how cosmological features of DM, particularly those linked to the reheating epoch, can be probed via missing energy-based searches at the ILC. Finally, we summarize our findings and conclude in Sec.~\ref{sec:conc}.

\section{Leptophilic DM operator under scrutiny}
\label{sec:model}
In absence of any experimental hint for NP, effective theory (EFT) offers a more judicious tool for such exploration. Among a broad class of effective theories that connect DM to the SM, the most relevant is the DMEFT framework~\cite{Duch:2014xda}:
\begin{equation}\label{eq:EFTop}
    \mathcal{L}_{\rm DMEFT} = \mathcal{L}_{\rm SM} + \mathcal{L}_{\rm DM} + \sum_{i,d} \frac{C_{i}^{(d)} \mathcal{O}^{(d)}_{i}}{\Lambda^{d-4}}\,,
\end{equation}
where $\mathcal{L}_{\rm SM}$ and $\mathcal{L}_{\rm DM}$ are the SM and DM Lagrangians, respectively, without any interaction between the two sectors. It should be noted that $\mathcal{L}_{\rm DM}$ in principle contains DM self-interaction terms, but for simplicity, we assume such interactions to be negligible to guide DM freeze out. The SM-DM interaction is mediated by $d$-dimensional EFT operators, $\mathcal{O}^{(d)}_{i}$, suppressed by the effective scale $\Lambda$, with $C_{i}^{(d)}$ being the dimensionless Wilson coefficients. Several works have cataloged subsets of DMEFT operators~\cite{Duch:2014xda,Criado:2021trs,Aebischer:2022wnl} and numerous studies have explored their implications in DM phenomenology, including search strategies.

A particularly compelling class of models features leptophilic operators \cite{Barman:2021hhg,Borah:2024twm,Ge:2023wye,Kundu:2021cmo}, where DM couples directly to the SM leptons. One class of such operators are current current interactions like,
\begin{equation}
    \mathcal{O}: J^{\rm{SM}}J_{\rm{DM}}\sim(\bar{\ell}\Gamma{\ell})(\phi \Gamma \phi)\,,
\end{equation}
where $\ell$ generically indicate left handed lepton doublet or right handed singlet, $\phi$ indiactes DM particle, and $\Gamma=\{1,\gamma^\mu,\gamma^5, \gamma^{\mu}\gamma^5, \sigma^{\mu\nu}\}$. While the LHC has limited sensitivity to such interactions due to suppressed production rates and overwhelming SM backgrounds, future high-energy lepton colliders offer a promising platform to probe these operators with high precision. Direct search via nuclear recoil for such DM operators also occur in one loop \cite{Barman:2021hhg}, resulting in reduced limit in both DM mass and NP scale. Electron recoil is naturally more sensitive to such leptophilic operators particularly with low DM mass. Unfortunately, such operators are not collider friendly, as the mono-X signal they produce, via initial state radiation of X ($=\gamma, Z$), do not see the Lorentz structure via which DM couples to SM. They produce a very similar missing energy pattern to that of SM background contamination, providing very little possibility to discover such DM. Therefore, we need to look for operators where the visible X produced, won't appear from the initial state radiation, but from the vertex itself. Some such operators that have been studied before \cite{Barman:2024nhr,Barman:2024tjt} involve FIMP coupled to the field strength tensors like $B_{\mu\nu}B^{\mu\nu}$ or $W_{\mu\nu}W^{\mu\nu}$, which provides distinguishable mono-$\gamma$ or mono-$Z$ signal at electron-positron collider. However, WIMPs that couple to such operators are heavily constrained by the indirect search limits.       

In this work, we focus on a gauge-singlet scalar DM $\phi$, whose interactions with SM leptons are described by the following dimension-6 effective operator:
\begin{equation}\label{eq:EFToptour}
\mathcal{O}_{\rm DM-SM}:~\sum_{i=1}^{3} \frac{(\overline{\ell}_{L}^{\,i} H e_{R}^{\,i})\, \phi^2}{\Lambda^{2}}\,,
\end{equation}
where $\ell_L^{\,i}$ and $e_R^{\,i}$ denote the SM left-handed lepton doublet and right-handed charged lepton singlet of the $i$-th generation ($i = 1, 2, 3$), and $H$ is the SM Higgs doublet. This operator allows scalar dark matter $\phi$ to couple to all three SM lepton families via Yukawa-like interactions. We assume a $\mathcal{Z}_2$ symmetry under which $\phi \to -\phi$, which stabilises DM, and allows terms in powers of $\sim \phi^2$. Since the combination $C_{i}/\Lambda^2$ governs the phenomenology, we fix the Wilson coefficient $C_{i} = 1$ for each lepton generation, to reduce the model parameters without loss of generality. The operator structure mirrors that of the SM Yukawa interaction and, after electroweak symmetry breaking (EWSB), leads to a 4-point vertex involving DM and SM leptons. This interaction facilitates DM production in the early Universe and contributes to the thermal population of the dark sector. In this study, we explore the phenomenological implications of this operator in both the standard freeze-out scenario, where DM decouples after the completion of the reheating phase, and a non-standard cosmological setting in which DM freeze-out occurs during the reheating epoch. Furthermore, the same operator gives rise to a five-point interaction involving DM, SM leptons, and the Higgs boson, leading to distinctive collider signatures at the lepton colliders. Thus, the model provides a unified framework where the same interaction controls both cosmological and collider observables, so that a signal discovery can hint towards the cosmological interpretation.  

\section{DM Genesis in the Early Universe}
\label{sec:dmprod}
As stated previously, our focus lies on scenarios in which DM is thermally produced through the freeze-out mechanism, commonly known as WIMPs. WIMPs initially remain in thermal equilibrium due to their sizable interactions with the thermal bath particles. However, as the Universe expands, the interaction rate eventually drops below the Hubble expansion rate, which ultimately helps DM to decouple or freeze-out from the bath. In this work, we investigate the freeze-out production of scalar DM, with an interaction term following Eq.~\ref{eq:EFToptour}, during reheating following the inflationary phase of the universe. 

\subsection{Post-Inflationary Dynamics}
\label{sec:reheat}
To explore the DM phenomenology in the reheating epoch, it is better to take a look into the post-inflationary evolution of the inflation field, whose decay gives rise to the radiation bath that facilitates the production of DM. After the end of inflation, the inflaton field $\Phi$ oscillates near the minima of its potential $V(\Phi)$, which we assume to be 
\begin{equation}\label{eq:inpot}
    V(\Phi)=\lambda\frac{\Phi^n}{\Lambda_{\Phi}^{n-4}}\,,
\end{equation}
where $\lambda$ is a dimensionless coupling constant and $\Lambda_{\Phi}\leq10^{16}$ GeV~\cite{Planck:2018jri} represents the characteristic energy scale of inflation. It may be noted that the NP responsible for generating $\Lambda_{\Phi}$ is distinct from the scale $\Lambda$ that governs the DM-SM interactions, so the two can differ significantly. The effective inflaton mass $m_{\Phi}(a)$ can be derived from the second derivative of Eq.~\ref{eq:inpot} and is given by
\begin{equation}\label{eq:mphi}
    m_{\Phi}(a)^{2}=n(n-1)\lambda\frac{\Phi^{n-2}}{\Lambda^{n-4}_{\Phi}}\simeq n(n-1)\lambda^{\frac{2}{n}}\Lambda_{\Phi}^{\frac{2(4-n)}{n}}\rho^{}_{\Phi}(a)^{\frac{(n-2)}{n}}\,,
\end{equation}
where $a$ is the scale factor of the universe. When $n\neq2$, $m_{\Phi}$ acquires a field-dependent form, which consequently creates a time-dependent decay rate for the inflaton field. The equation of motion of the oscillating inflaton field can be written as 
\begin{equation}\label{eq:eqm}
    \ddot \Phi+(3\mathcal{H}+\Gamma_{\Phi})\dot\Phi+V^{\prime}(\Phi)=0\,.
\end{equation}
Here, $\Gamma_{\Phi}$ represents the inflaton decay rate, $\mathcal{H}$ denotes the Hubble expansion rate of the universe, and the overdot ($\dot{}$) and prime ($\prime$) indicate derivatives with respect to time $t$ and the field $\Phi$, respectively. The evolution of the inflation energy density, $\rho^{}_{\Phi}$ is governed by the following Boltzmann equation (BEQ):
\begin{equation}\label{eq:phieom}
    \frac{d\rho^{}_{\Phi}}{dt}+\frac{6n}{2+n}\mathcal{H}\rho_{\Phi}^{}=-\frac{2n}{2+n}\Gamma_{\Phi}\rho_{\Phi}^{}\,,
\end{equation}
where $\mathcal{H}=\sqrt{(\rho^{}_{R}+\rho_{\Phi}^{})/M_{Pl}^{}}$ is the expansion rate of the universe and the reduced Planck mass is denoted as $M_{Pl}^{}=2.435\times10^{18}$ GeV. The oscillating inflaton field can be described as a fluid with an equation of state $\omega\equiv p_{\Phi}^{}/\rho_{\Phi}^{}=(n-2)/(n+2)$~\cite{Turner:1983he}, where the inflaton pressure is given by $p_{\Phi}^{}=\frac{1}{2}\dot\Phi^{2}-V(\Phi)$. In Eq.~\ref{eq:phieom}, the $\Gamma_{\Phi}^{}\rho_{\Phi}^{}$ is responsible for the transfer of energy from the inflaton field to the radiation bath through its decay, while $\mathcal{H}\rho_{\Phi}^{}$ accounts for the redshift dilution due to the expansion of the Universe. During the reheating phase of the Universe, when $a_{\rm I}^{}\ll a\ll a_{\rm rh}^{}$, the expansion term dominates over the decay term; here $a_{I}$ is the scale factor of the universe at the beginning of reheating, and $a_{\rm rh}$ represents that after completion of reheating. This enables us to find an analytical solution for $\rho_{\Phi}^{}$ of the form
\begin{equation}\label{eq:phian}
    \rho_{\Phi}^{}(a)\simeq\rho_{\Phi}^{}(a_{\rm rh}^{})\Big(\frac{a_{\rm rh}^{}}{a}\Big)^{\frac{6n}{2+n}}\,.
\end{equation}
 As the Hubble expansion rate during reheating is primarily governed by the inflation energy density, Eq.~\ref{eq:phian} enables us to write the expression for Hubble expansion as
\begin{equation}\label{eq:Hubble}
\mathcal{H}(a) \simeq \mathcal{H}(a_{\rm rh}) \times \begin{cases}
\left(\frac{a_{\rm rh}}{a}\right)^{\frac{3n}{(n + 2)}} &\text{for } a \leq a_{\rm rh},\\[10pt]
\left(\frac{a_{\rm rh}}{a}\right)^2 &\text{for } a_{\rm rh} \leq a.
\end{cases}
\end{equation}
The BEQ that describes the SM radiation energy density $\rho_{R}^{}$, can be written as~\cite{Garcia:2020wiy} 
\begin{equation}\label{eq:Reom}
    \frac{d\rho^{}_{R}}{dt}+4\mathcal{H}\rho_{R}^{}=+\frac{2n}{2+n}\Gamma_{\Phi}\rho_{\Phi}^{}\,.
\end{equation}
The reheating phase ends when the radiation and inflation energy densities become equal, i.e.,
\begin{equation}\label{eq:matradeq}
    \rho_{R}^{}(a_{\rm rh}^{})=\rho_{\Phi}^{}(a_{\rm rh}^{})=3M_{Pl}^{2}\mathcal{H}(a_{\rm rh}^{})^{2}\,.
\end{equation}
To respect the successful prediction of BBN, the reheating temperature $T_{\rm rh}^{}$ must exceed the BBN temperature, $T_{\rm BBN}^{}\simeq4$ MeV ($T_{\rm rh}^{}>T_{\rm BBN}^{}$)~\cite{Sarkar:1995dd, Kawasaki:2000en, Hannestad:2004px, deSalas:2015glj, Hasegawa:2019jsa, DeBernardis:2008zz}.
\subsection{Reheating via perturbative process}

Here we briefly discuss the perturbative process of reheating~\cite{Abbott:1982hn} from inflaton decay, where the interaction and decay rate of inflaton particles are calculated using perturbative coupling expansions. For two-body decay channels, we consider a simple Lagrangian of the inflaton coupling with the SM as,
\begin{equation}\label{eq:inflag}
    \mathcal{L}\supset -\mu_{\chi}\Phi|\chi|^2-y_{\Psi}^{}\Phi\bar{\Psi}\Psi\,,
\end{equation}
where $\chi$ can be a complex scalar doublet (such as the SM Higgs doublet) with dimensionful coupling $\mu_{\chi}$ to interact with the inflaton. $\Psi$ can be a fermion belonging to thermal bath with coupling $y_{\Psi}^{}$ to interact with $\Phi$. While writing the interactions of Eq.~\ref{eq:inflag}, we assume that the inflaton decays exclusively to the visible sector particles, which then interact and thermalise among themselves, forming the radiation bath. In the analysis that follows, each reheating scenario (fermionic and bosonic) is considered separately, assuming that both scenarios cannot take place concurrently.
\subsubsection{Reheating via Fermionic Channel}
\label{sec:reheatfer}
In this section, we discuss the case of fermionic reheating, where the inflaton field decays solely into the vector-like fermion pair ($\bar{\Psi}\Psi$) via the Yukawa interaction as in Eq.~\ref{eq:inflag}; the associated decay rate is given by:
\begin{equation}
    \Gamma_{\Psi}=\frac{y_{\rm eff}^{2}}{8\pi}m_{\Phi}^{}(a)\,,
\end{equation}
where the effective coupling $y_{\rm eff}\neq y_{\Psi}^{}$ (for $n\neq 2$) is derived by averaging over multiple oscillations~\cite{Garcia:2020wiy, Shtanov:1994ce, Ichikawa:2008ne}. By solving Eq.~\ref{eq:Reom}, one can find the following expression~\cite{Bernal:2022wck, Barman:2023rpg, Xu:2024fjl} for the radiation energy density in terms of scale factor $a$:
 \begin{equation}\label{eq:rhoRfer}
     \rho_{R}^{}(a)\simeq\frac{\sqrt{3}}{8\pi}\frac{n\sqrt{n(n-1)}}{7-n}\frac{y_{\rm eff}\lambda^{\frac{1}{n}}M_{Pl}}{\Lambda_{\Phi}^{\frac{n-4}{n}}\rho_{\Phi}^{}(a_{\rm rh})^{\frac{1-n}{n }}}\Big(\frac{a_{\rm rh}}{a}\Big)^{\frac{6(n-1)}{2+n}}\Bigg[1-\Big(\frac{a_{\rm I}^{}}{a}\Big)^{\frac{2(7-n)}{2+n}}\Bigg]\,.
 \end{equation}
 The temperature of the thermal bath is given by:
 \begin{equation}
T(a) \simeq T_{\rm rh}\times \left(\frac{a_{\rm rh}}{a}\right)^\beta,
\label{eq:Tevol}
\end{equation}
with
\begin{equation}
\beta =
\begin{cases}
\frac32\, \frac{n - 1}{n + 2} & \text{ for } n < 7\,,\\
1 & \text{ for } n > 7\,.
\end{cases}
\label{eq:Tfer}
\end{equation}
The Hubble expansion rate during reheating can be expressed in terms of temperature as
\begin{equation}
\mathcal{H}(T) \simeq \mathcal{H}(T_{\rm rh}) \left( \dfrac{T}{T_{\rm rh}} \right)^{\frac{3}{\beta}\frac{n}{n+2 } \,} \,,
\label{eq:Hevol}
\end{equation}
where $T_{\rm rh}$ denotes the reheat temperature, marking the end of the reheating phase. 
\begin{figure}[h]
    \centering
    \includegraphics[width=0.49\linewidth]{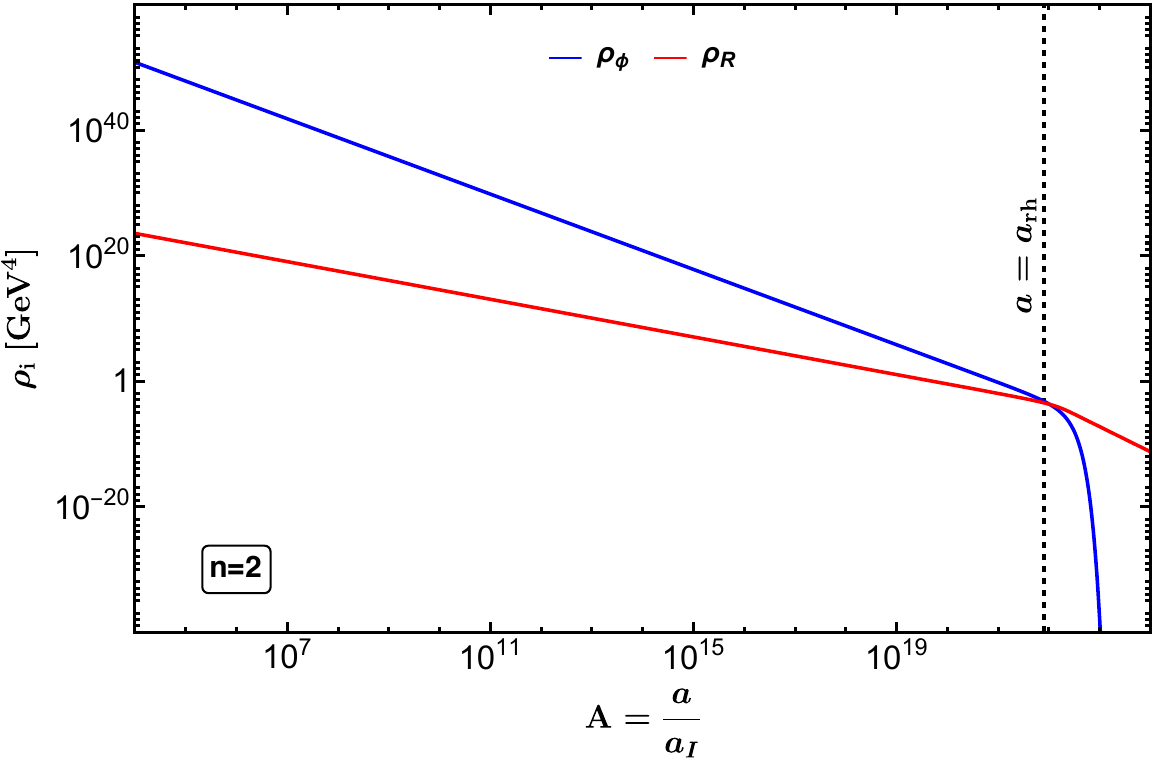}
     \includegraphics[width=0.49\linewidth]{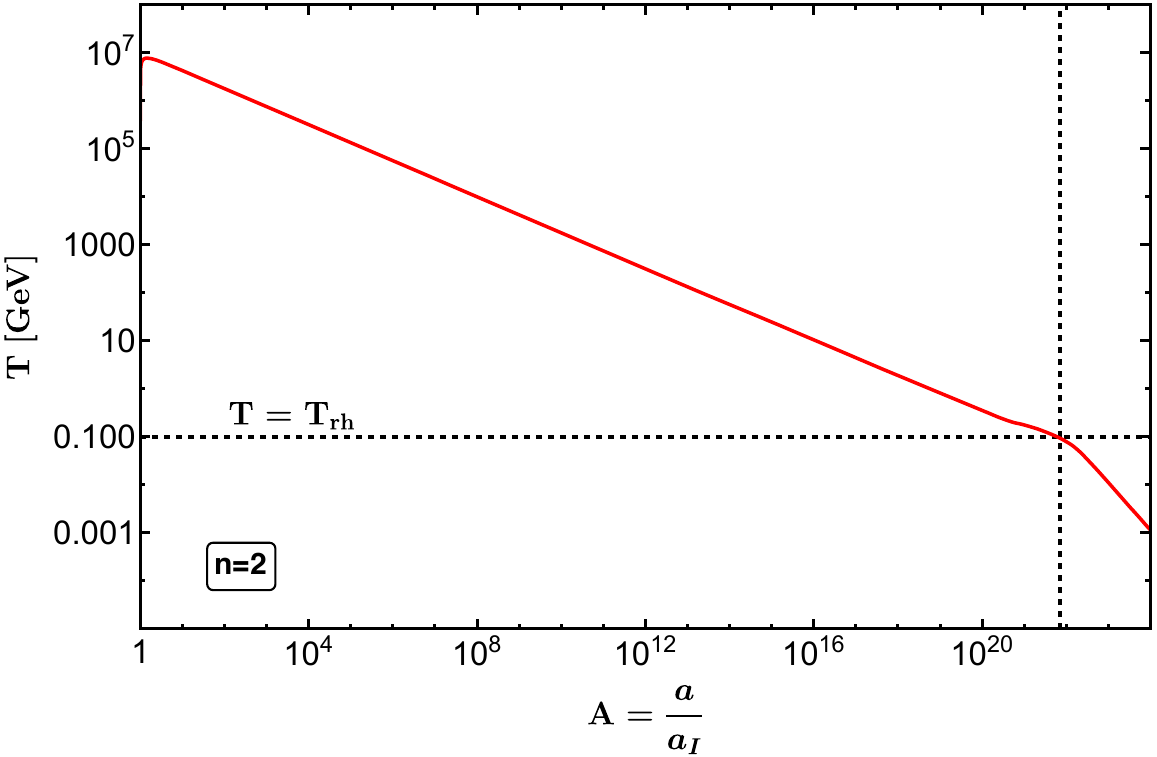}
     \includegraphics[width=0.49\linewidth]{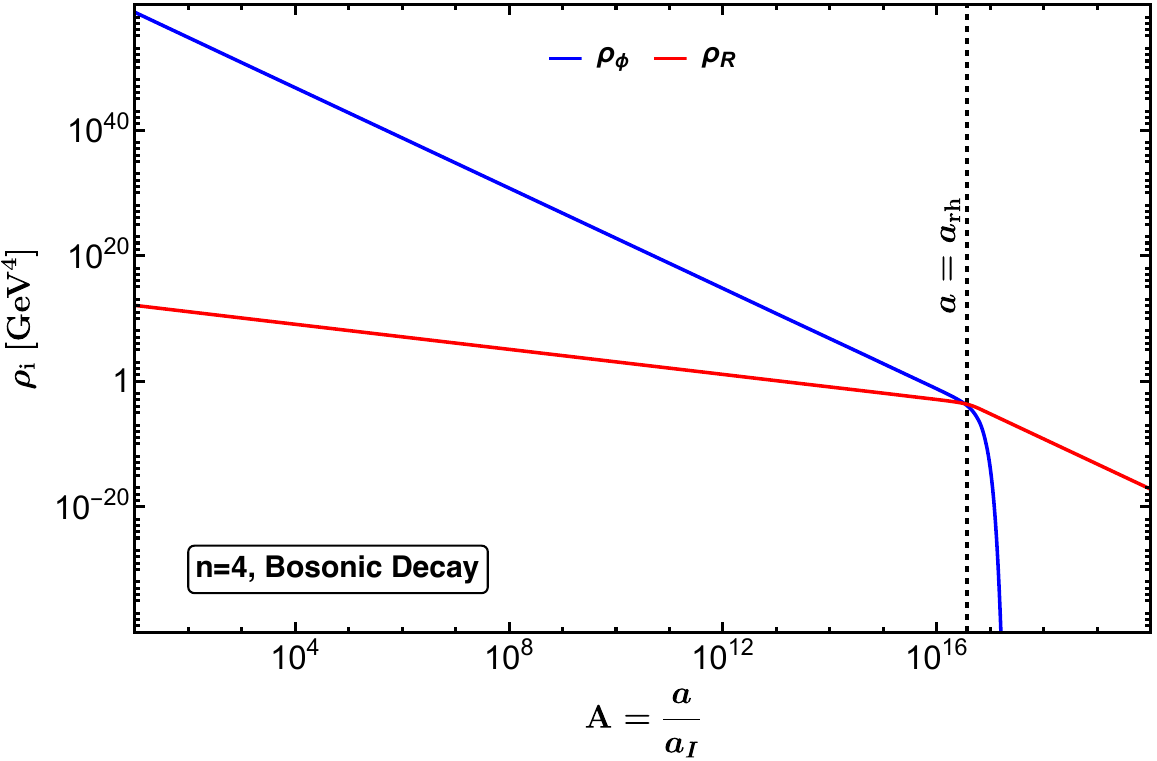}
     \includegraphics[width=0.49\linewidth]{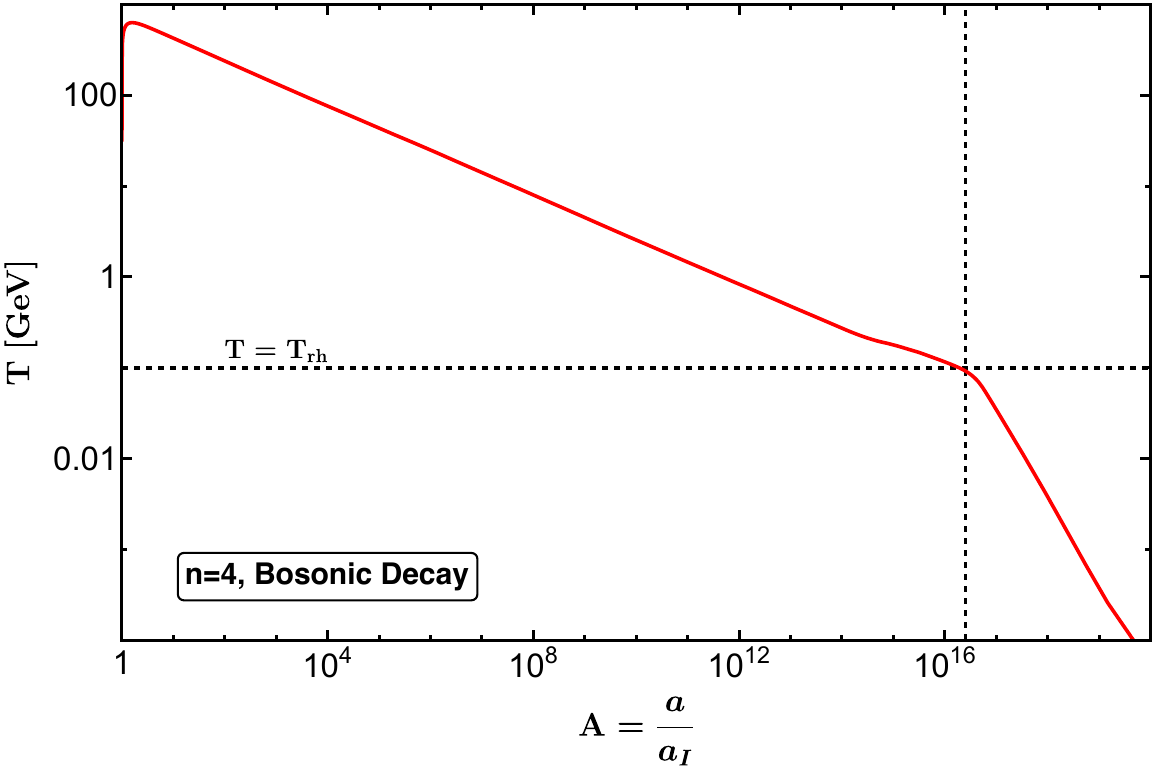}
     \includegraphics[width=0.49\linewidth]{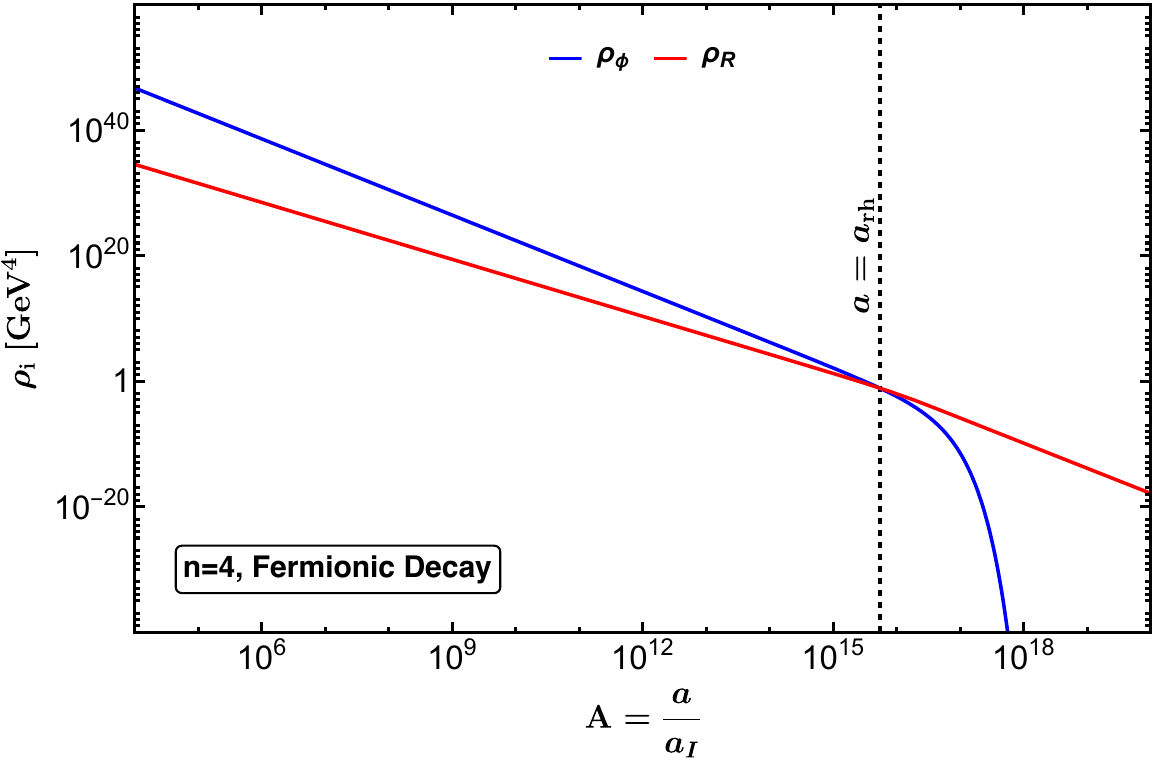}
     \includegraphics[width=0.49\linewidth]{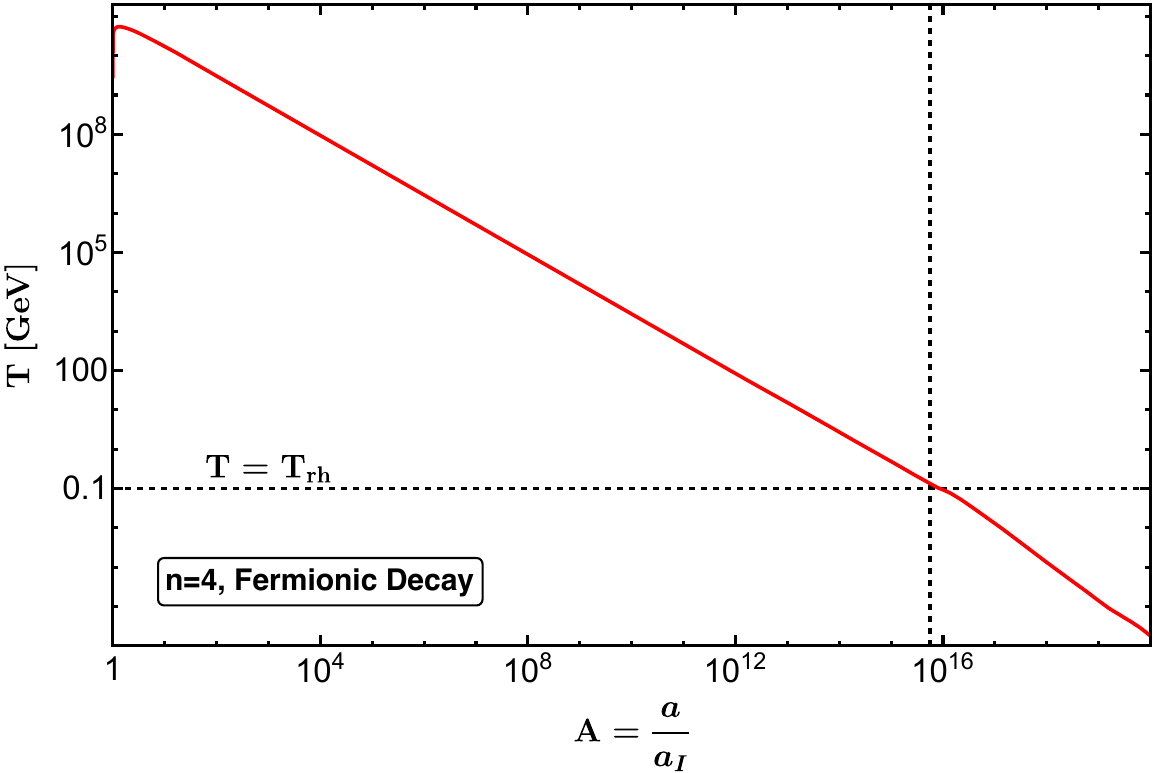}
    \caption{Evolution of inflation and radiation energy densities ( left panel) and corresponding thermal bath temperature (right panel) as a function of scale factor $\rm a$. Results are shown for both bosonic and fermionic reheating scenarios with representative values of $n$. The vertical dashed line ($a=a_{\rm rh}^{}$) indicates the onset of the radiation domination epoch.}
    \label{fig:rhoT}
\end{figure}
\subsubsection{Reheating from Bosonic Channel}
\label{sec:reheatbos}
When the inflaton exclusively decays into a pair of bosons through the trilinear interaction (see Eq.~\ref{eq:inflag}) the corresponding decay rate is:
\begin{equation}
    \Gamma_{\chi}=\frac{\mu_{\rm eff}^{2}}{8\pi~m_{\Phi}^{}(a)}\,.
\end{equation}
Similar to the fermionic case, the effective coupling $\mu_{\rm eff}\neq\mu$ (for $n\neq2$) can be derived by averaging over multiple oscillations. Following a similar approach as before, the radiation energy density for bosonic reheating evolves as~\cite{Bernal:2022wck}:
\begin{equation}\label{eq:rhoRbos}
    \rho_{R}^{}(a)\simeq \frac{\sqrt{3}}{8\pi}\frac{1}{1+2n}\sqrt{\frac{n}{n-1}}\frac{\mu_{\rm eff}^{2}M_{Pl}}{\lambda_{\Phi}^{\frac{1}{n}}\Lambda^{\frac{4-n}{n}}}\rho_{\Phi}^{}(a_{\rm rh})\Big(\frac{a_{\rm rh}}{a}\Big)^{\frac{6}{2+n}}\Bigg[1-\Big(\frac{a_{\rm I}^{}}{a}\Big)^{\frac{2(2n+1)}{2+n}}\Bigg]\,.
\end{equation}
For the bosonic case also, during reheating the SM bath temperature and the Hubble expansion evolve as given by Eq.~\ref{eq:Tevol} and Eq.~\ref{eq:Hevol}, respectively, with $\beta=\frac{3}{2}\frac{1}{n+2}$. The maximum temperature attained by the thermal bath during reheating is approximately given by:
\begin{align}
T_{\rm max} \simeq T_{\rm rh} \times
\begin{cases}
\left(\dfrac{a_{\rm rh}}{a_{\rm I}^{}}\right)^{\frac{3(n - 1)}{2(n + 2)}} & \text{fermionic reheating}, \\[10pt]
\left(\dfrac{a_{\rm rh}}{a_{\rm I}^{}}\right)^{\frac{3}{2(n + 2)}} & \text{bosonic reheating},
\end{cases}
\end{align}
which can exceed $T_{\rm rh}$ by several orders of magnitude when $a_{\rm rh}\gg a_{\rm I}^{}$. Left panels of Fig.~\ref {fig:rhoT} show the evolution of inflation (blue) and radiation (red) energy densities as a function of the scale factor for $T_{\rm rh}=100$ MeV (for both fermionic and bosonic reheating scenarios). The right panels display the evolution of the SM bath temperature. The small bump around $150-180$ MeV~\cite{Schwarz:2003du} corresponds to the QCD phase transition, which causes a rapid drop in the effective number of relativistic degrees of freedom. The top panels illustrate the case $n=2$, where the fermionic and bosonic decay processes produce identical dynamics. During reheating, in this case, the energy densities and temperature scales $\rho_{\Phi}^{}\propto a^{-3}$, $\rho_{R}^{}\propto a^{-3/2}$, and $T\propto a^{-3/8}$. The middle and bottom panels correspond to $n=4$, with bosonic and fermionic reheating scenarios, respectively. In the bosonic case, $\rho_{R}^{}\propto a^{-1}$ and $T\propto a^{-1/4}$, while in the fermionic case, $\rho_{R}^{}\propto a^{-3}$ and $T\propto a^{-3/4}$. As a result, for $n > 2$, reheating proceeds more efficiently over time when the inflaton decays into bosonic final states compared to fermionic states. 
\begin{figure}[htb!]
	\centering
	\includegraphics[width=0.65\linewidth]{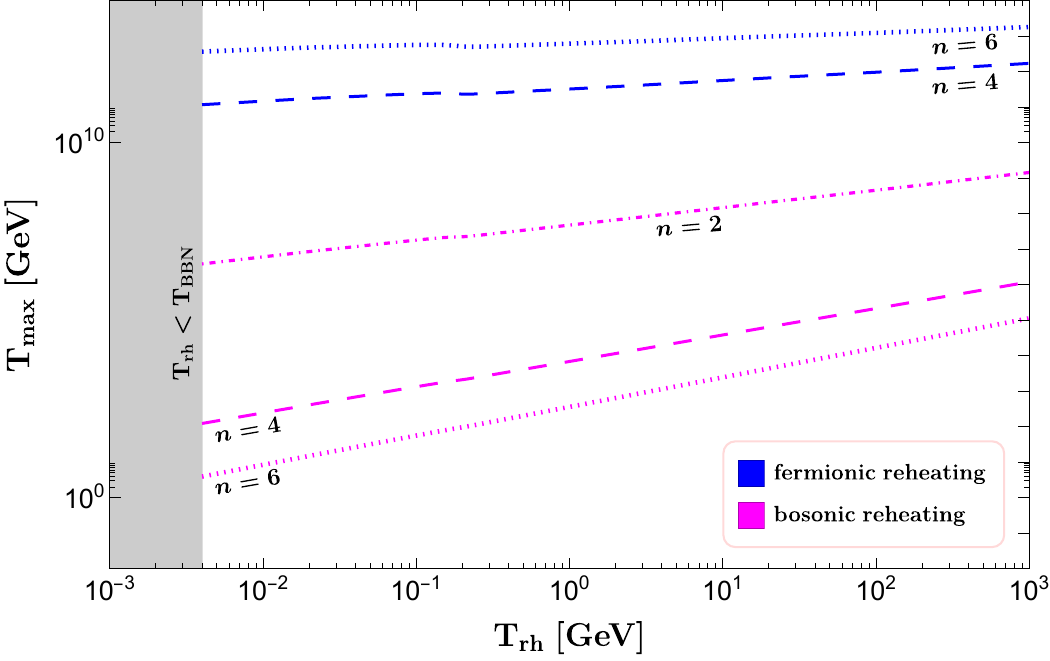}
	\caption{The maximum temperature of the bath attained during reheating, $\rm T_{max}$, plotted as a function of the reheating temperature, $\rm T_{rh}$. The dot-dashed, dashed and dotted lines represent the cases with $n=2$, $n=4$ and $n=6$, respectively. The dark shaded region is excluded from BBN as in this region $\rm T_{rh}<T_{BBN}$. For $n=2$, the fermionic and bosonic reheating curves coincide, represented by the dot-dashed line.}
	\label{fig:tmaxtrh}
\end{figure}
Fig.~\ref{fig:tmaxtrh} shows that in the case of fermionic reheating, the maximum bath temperature $T_{\rm max}$ exceeds $\mathcal{O}(10^4)$ GeV for all values of $n$. Since our goal is to investigate the reheating scenarios in a collider with $\sqrt{s}=1$ TeV, we focus on cases where $\Lambda\gtrsim1$ TeV. However, the effective description must respect $\Lambda>T_{\rm max}$, this condition ensures that the effective description of interaction is valid during the reheating phase. Abiding by this restriction, we observe that the associated effective NP scale for the fermionic reheating lies far beyond the reach of current collider experiments. In contrast, the bosonic reheating scenario, particularly for $n=4$ and $n=6$ allows a lower $T_{\rm max}$, and thus ensures a more accessible $\Lambda$. This motivates us to focus exclusively on the bosonic reheating scenario for our DM phenomenological study.

Before proceeding to the next section, we briefly comment on the preheating effect. Lattice simulations suggest that during reheating for $n\gtrsim3$, inflation fragmentation and parametric resonance~\cite{Lozanov:2016hid, Maity:2018qhi, Saha:2020bis, Antusch:2020iyq} can drive the equation of state parameter $\omega\rightarrow1/3$, effectively pushing the Universe towards radiation domination. However, complete depletion of inflation energy still requires perturbative decays via trilinear couplings at the final stage of reheating~\cite{Maity:2018qhi}. Efficient inflation energy depletion via preheating typically demands large couplings, which can lead to significant loop corrections, potentially distorting the inflation potential and altering inflationary predictions. Moreover, studies~\cite{Easther:2010mr} have shown that coherent oscillations of the inflaton field may break down, causing delayed preheating. Additionally, depending on the spin of the decay products of the inflaton, preheating can be suppressed by a large effective Higgs vacuum expectation value (VEV, $v_{\rm EW}^{}$)~\cite{Freese:2017ace}. Since we are interested in low-temperature reheating with small inflaton couplings, preheating is expected to be insufficient. Hence, for the remainder of this work, we carry out our analysis assuming a perturbative reheating framework. 

With an understanding of the background cosmological evolution, we will study the DM phenomenology in the next subsection, with particular attention to the case where freeze-out occurs during the reheating epoch.

\subsection{Production Mechanism of Thermal Dark Matter}

The evolution of the DM number density $n_{\phi}^{}$ can be determined by solving the following BEQ 
\begin{equation}\label{eq:BEQDM}
    \frac{dn_{\phi}^{}}{dt}+3\mathcal{H}n_{\phi}^{}=-\langle\sigma v\rangle(n^{2}_{\phi}-n^{2}_{\rm eq})\,.
\end{equation}
Here, $n_{\rm eq}^{}\simeq g(m_{\phi}^{}~T/2\pi)^{3/2}e^{-m_{\phi}^{}/T}$ is the equilibrium number density of the DM ($\phi$), where $g$ and $m_{\phi}^{}$ are its number of degrees of freedom and mass, respectively. The quantity $\langle\sigma v\rangle$ represents the thermal average cross section times relative velocity, defined as~\cite{Gondolo:1990dk}:
\begin{equation}\label{eq:sigv}
    \langle\sigma v\rangle=\frac{g^2~T}{32\pi^4~n_{eq}^{2}}\int_{4m_{\phi}^{2}}^{\infty}\sqrt{s}(s-4m_{\phi}^{2})\sigma(s)K_{1}\Bigg(\frac{\sqrt{s}}{T}\Bigg)~ds\,,
\end{equation}
where $\sigma(s)$ is the DM annihilation cross-section to SM particles. 
\begin{figure}[htb]
	\centering
	\begin{tikzpicture}[baseline={(current bounding box.center)},style={scale=1, transform shape}]
	\begin{feynman}
		\vertex[square dot](a){};
		\vertex[above left =1.25cm and 2cm of a] (a1){$\phi$};
		\vertex[below left =1.25cm and 2cm of a] (a2){$\phi$};
		\vertex[above right = 1.25cm and 2cm of a] (b1) {$\ell^{-}$};
		\vertex[below right = 1.25cm and 2cm of a] (b2){$\ell^{+}$};
		\diagram*{
			(a) -- [thick, scalar, arrow size=1pt] (a2),
			(a1) -- [thick, scalar, arrow size=1pt] (a),
			(a) -- [thick, fermion, arrow size=0.7pt] (b1),
			(b2) -- [thick, fermion, arrow size=0.7pt] (a)
		};
	\end{feynman}
	\end{tikzpicture}
	\caption{Feynman diagrams corresponding to the DM relic density. Here, $\ell = e, \mu, \tau$. The square dot corresponds to the EFT vertex.}
    \label{fig:dm0}
\end{figure}
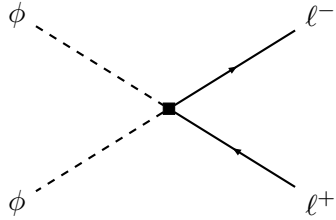
In our case (see Eq.~\ref{eq:EFToptour}), this cross-section is given by: 
\begin{equation}
	\sigma(s)\equiv\sigma_{\phi \phi \to \ell^{+} \ell^{-}}(s) = \frac{\beta_{\phi} v_{\rm EW}^{2}}{4\pi \Lambda^{4}} \frac{(s - 4 m^{2}_{\ell})}{(s - 4 m^{2}_{\phi})}\,,
    \label{ID_eq1}
\end{equation}
where, $\ell = e, \mu, \tau$, and,
\begin{equation}
	\beta_{\phi} =  \sqrt{\frac{(s - 4 m^{2}_{\ell})(s - 4 m^{2}_{\phi})}{s^{2}}} \,.
\end{equation}
From Eqs.~\ref{ID_eq1} and \ref{eq:sigv}, one can find the following analytical expression for $\langle\sigma_{\phi \phi \to \ell^{+} \ell^{-}}v\rangle$
\begin{equation}\label{eq:sigmavan}
    \langle\sigma_{\phi \phi \to \ell^{+} \ell^{-}}v\rangle=\frac{g^2}{n_{\rm eq}^{2}}\times\frac{v_{\rm EW}^{2}~T^6}{4\pi^5~\Lambda^{4}}\times\mathcal{I}(m_{\phi},T)\,,
\end{equation}
where 
\begin{align}
\mathcal{I}(m_{\phi},T) \approx
\begin{cases}
\frac{\sqrt{2\pi}}{32}\Big(\frac{2 m_{\phi}}{T}\Big)^{7/2} e^{-2m_{\phi}/T}\left[1+\frac{7}{2}\frac{1}{\left(\frac{2m_{\phi}}{T}\right)}+\frac{35}{4}\frac{1}{\left(\frac{2m_{\phi}}{T}\right)^2}+\dots\right] & \text{for}~m_{\phi}> T, \\[20pt]
\left[1-\frac{1}{4}\left(\frac{m_{\phi}}{T}\right)^2+\frac{1}{8}\left(\frac{m_{\phi}}{T}\right)^4\Big\{\gamma+\log\big[\frac{m_{\phi}}{T}\big]\Big\}+\mathcal{O}\left(\left(\frac{m_{\phi}}{T}\right)^6\right)\right] & \text{for}~m_{\phi}< T.
\end{cases}
\end{align}
In Eq.~\ref{eq:sigmavan}, $T$ represents the DM temperature, which can be assumed to be the bath temperature as the DM was in equilibrium before freeze out. Here, $\gamma \approx 0.577215$ denotes the Euler--Mascheroni constant. It is worth mentioning that DM can also be produced directly from inflaton decays. In such cases, we should include an additional source term proportional to $\mathcal{B}_{\rm DM}\Gamma_{\Phi} n_{\Phi}^{}$ in Eq.~\ref{eq:BEQDM}, where $\mathcal{B}_{\rm DM}$ is the branching ratio for inflaton decay into DM pairs, and $n_{\Phi}^{}$ is the inflaton number density. However, since our focus is on the impact of the time-varying decay rate of the inflaton on the production of thermal WIMP, we neglect the effect of DM production directly from the inflaton decays in our current analysis. For a detailed discussion on DM production during the thermalization phase, we refer the reader to the relevant literature~\cite {Harigaya:2014waa, Harigaya:2019tzu, Drees:2021lbm, Drees:2022vvn, Mukaida:2022bbo}.

\subsubsection{Freeze-Out after Reheating}
\label{sec:fo-arh}
Here we quickly revisit the conventional freeze-out scenario, where DM decouples from the thermal bath well after the reheating phase, during the radiation-dominated (RD) era. Throughout this RD regime, SM entropy remains conserved, and Eq.~\ref{eq:BEQDM} can be reformulated in a more convenient form in terms of the dimensionless quantity $x\equiv m_{\phi}/T$ and DM yield $Y\equiv n_{\phi}/s$ as
\begin{equation}\label{eq:BEQDMYx}
    \frac{dY}{dx}=-\frac{s~\langle\sigma v\rangle}{\mathcal{H}~x}\left(Y^2-Y^2_{\rm eq}\right)\,,
\end{equation}
where $s(T)=\frac{2\pi^2}{25}g_{s}T^3$ is the SM entropy density with $g_{s}(T)$ being the effective degrees of freedom of the thermal bath associated with entropy. Unlike the reheating phase, in the RD era, the Hubble parameter is given by $H(T)=\sqrt{\frac{\rho_{R}^{}}{3 M_{Pl}^2}}=\sqrt{\frac{\pi^2g_{\rho}^{}(T)}{90}}\frac{T^2}{M_{Pl}}$, where $g_{\rho}^{}(T)$ denotes the effective relativistic degrees of freedom associated with thermal bath energy density. From Eq.~\ref{eq:BEQDMYx}, after integrating,  one can find the approximated analytical form (for $s-$wave annihilation) of the present-time DM yield as
\begin{equation}\label{eq:DMyield}
    Y_{0}\simeq\frac{15}{2\pi\sqrt{10}}\frac{\sqrt{g_{\rho}^{}}}{g_{s}^{}}\frac{1}{M_{Pl}^{}T_{\rm fo}^{}\langle\sigma v\rangle}\,.
\end{equation}
Using Eq.~\ref{eq:DMyield}, one can find the following expression for the present DM relic abundance
\begin{equation}\label{eq:DMrelic}
    \Omega_{\rm DM}^{}h^2=\frac{m_{\phi}~Y_{0}~s_{0}^{}}{\rho_{c}^{}}h^2\propto m_{\phi}^{}~\Lambda^4\,,
\end{equation}
where $\rho_{c}^{}\approx10^{-5}h^2$ $\text{GeV}/\text{cm}^3$ is the critical energy density of the universe and $s_{0}^{}\approx2.7\times10^{3}~\text{cm}^{-3}$ is the present day entropy density of the universe. The freeze-out temperature is determined using the following condition $n_{\rm eq}^{}(T_{\rm fo})\langle\sigma v\rangle=\mathcal{H}(T_{\rm fo})$, and can be found out as:
\begin{equation}\label{eq:TfoRD}
    T_{\rm fo}=-\frac{2m_{\phi}}{\mathcal{W}_{-1}^{}\left[-\frac{8\pi^5}{45}\frac{g_{\rho}^{}}{g^2_{}}\frac{1}{\left(M_{Pl}^{}m_{\phi}^{}\langle\sigma v\rangle\right)^2}\right]}\,,
\end{equation}
where $\mathcal{W}_{-1}^{}$ represents the $-1$ brunch of the Lambert function.
\begin{figure}[htb!]
	\centering
	\includegraphics[width = 0.85\linewidth]{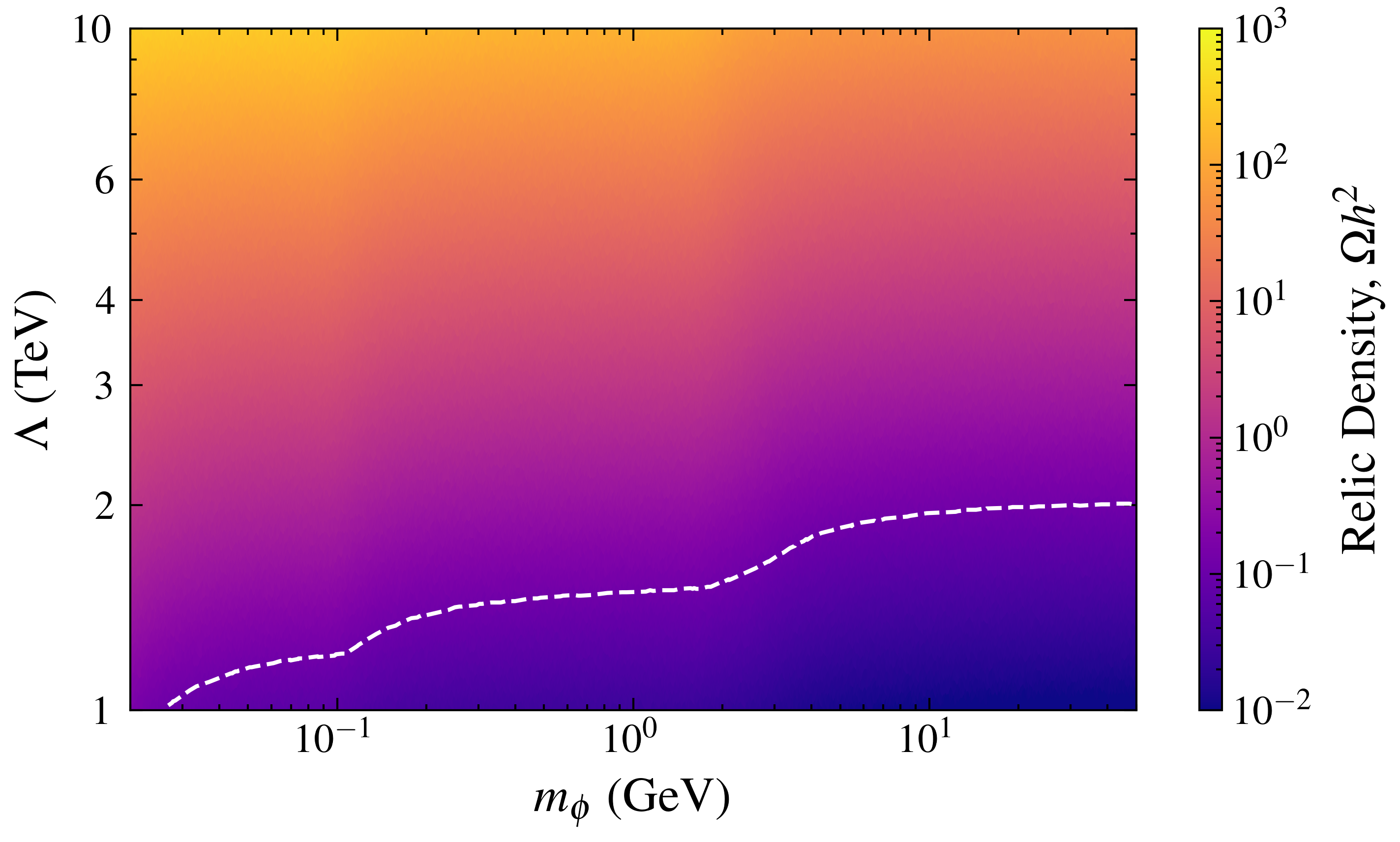}
	\caption{Relic-abundance satisfied line (\textit{white, dashed}) for scalar DM with freeze-out after reheating. The colour shades and colourbar correspond to DM relic density.}
	\label{fig:fo-after-rh}
\end{figure}
Fig.~\ref{fig:fo-after-rh} displays the DM relic density (indicated by colour gradient) corresponding to the leptophilic operator (see Eq.~\ref{eq:EFToptour}) in $\{m_{\phi}^{},~\Lambda\}$ parameter space using \texttt{MicrOMEGAs}~\cite{Alguero:2023zol} package. The white dashed contour represents the observed DM relic density, $\Omega_{\rm DM}^{}h^2\simeq0.12$ \cite{Planck:2018vyg}, with the two visible bumps along the curves corresponding to the departure from thermal equilibrium of $\tau$-lepton ($m_{\phi}^{}\sim 2$ GeV) and muon ($m_{\phi}^{}\sim 100$ MeV, along with QCD phase transition).
\subsubsection{Freeze-Out during Reheating}
\label{sec:fo-drh}
When freeze-out occurs during reheating, a period during which the SM entropy is not conserved due to the inflaton decay into the SM particles, it is more appropriate to express the BEQ of Eq.~\ref{eq:BEQDM} in the following form
\begin{equation}\label{eq:BEQDMN}
    \frac{dN}{d\rm A}=-\frac{\langle\sigma v\rangle}{\mathcal{H}\rm~A^4}\left(N-N^2_{\rm eq}\right)\,,
\end{equation}
where $\text{A}\equiv\frac{a}{a_{I}}$, $N\equiv n_{\phi}^{}\rm A^3$, and $N_{\rm eq}^{}\equiv n_{\rm eq}^{}\rm A^3$. After integrating and employing the Eq.~\ref{eq:Hubble}, one can find the following analytical solution of Eq.~\ref{eq:BEQDM}
\begin{equation}
    N({\rm A_{rh}^{}})\simeq\frac{6}{2+n}\frac{\mathcal{H}({\rm A_{rh}})}{\langle\sigma v\rangle}{\rm A_{rh}^{3}}\left(\frac{{\rm A_{rh}}}{{\rm A_{fo}}}\right)^{-\frac{6}{2+n}}\,,
\end{equation}
where ${\rm A_{fo}}$ and ${\rm A_{rh}}$ represent the values of the scale factor at the times when the temperature equals $\rm T_{fo}^{}$ and $\rm T_{rh}^{}$, respectively. Following the completion of the reheating epoch, that is, when $a>a_{\rm rh}^{}$, the thermal bath entropy becomes constant. As a result, the present-day DM yield $Y_{0}^{}$ for the case of bosonic reheating can be well approximated by its value at the end of reheating $Y(a_{\rm rh}^{})$, which can be quantified as~\cite{Bernal:2022wck}:
\begin{equation}
    Y_{0}^{}\equiv Y(a_{\rm rh}^{})=\frac{N({A_{\rm rh}^{}})}{s({A_{\rm rh}^{}}){\rm A_{rh}^{3}}}\simeq\frac{6}{2+n}\frac{45}{2\pi^2~g_{s}^{}}\frac{\mathcal{H}(T_{\rm rh}^{})}{\langle\sigma v\rangle~T_{\rm rh}^{3}}\left(\frac{T_{\rm rh}^{}}{T_{\rm fo}^{}}\right)^4\,.
\end{equation}
It is worth mentioning that in the period following freeze-out but preceding the completion of reheating, when $\rm T_{\rm rh}^{}\leq T\leq T_{\rm fo}^{}$, the DM yield evolves according to the relation $Y(T)\propto (a T)^{-3}$, which leads to the following expression for the bosonic reheating case
\begin{equation}
Y(T)\propto T^{2n+1}\,.    
\end{equation}
In the scenarios where DM freezes-out during reheating, the resulting DM relic density in the context of bosonic reheating exhibits an approximate dependence on the parameters $\{n,~m_{\phi}^{},~\rm T_{fo}^{},\rm T_{rh}^{}\}$, as given by:
\begin{equation}\label{eq:Omegarh}
    \Omega_{DM}^{}h^2\propto\frac{6}{2+n}\frac{m_{\phi}^{}\Lambda^4}{\rm T_{rh}^{}}\left(\frac{T_{\rm rh}^{}}{T_{\rm fo}^{}}\right)^4\,.
\end{equation}
 \begin{figure}[h]
    \centering
    \includegraphics[width=0.49\linewidth]{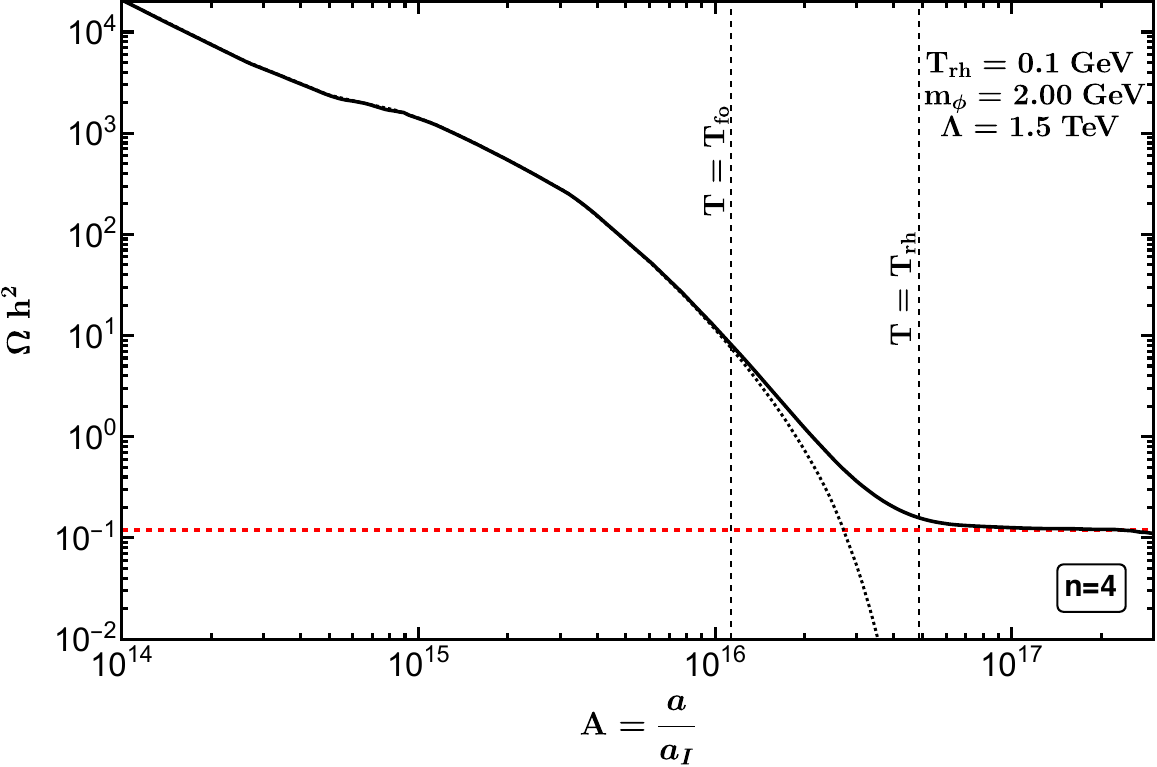}
     \includegraphics[width=0.49\linewidth]{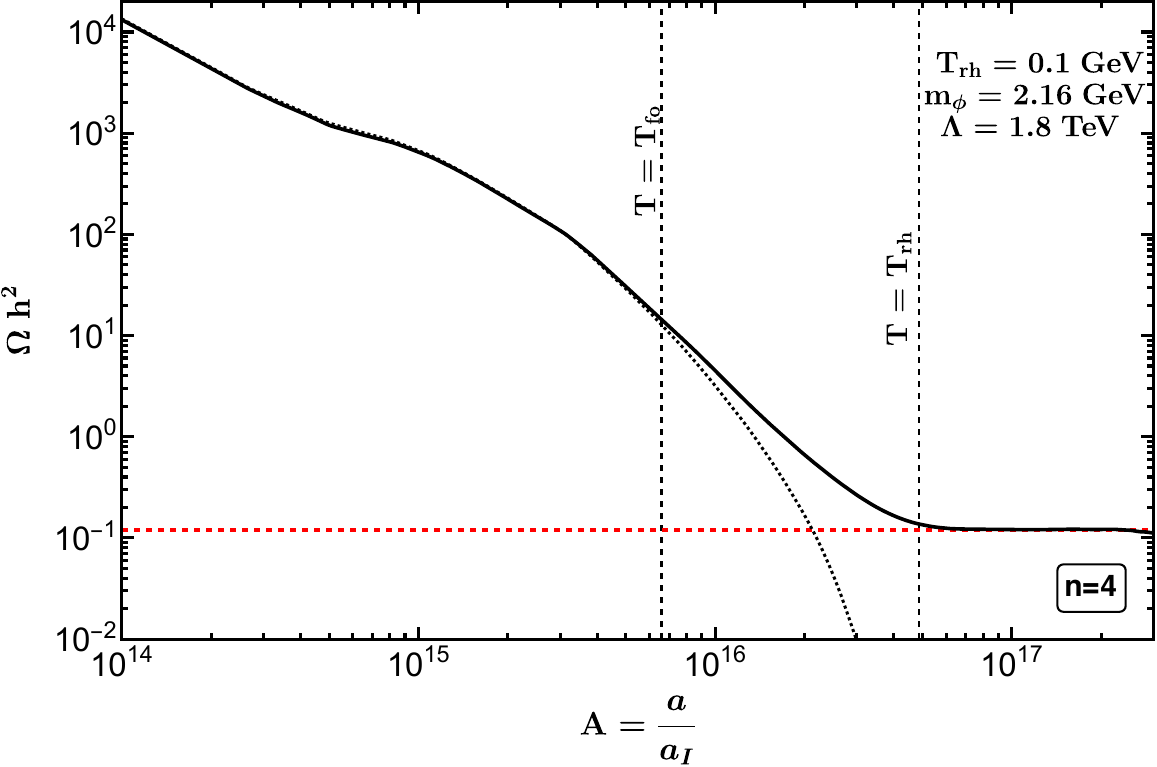}
    \caption{Evolution of DM relic density as a function of $\text{A}=\frac{a}{a_{I}}$ for two benchmark points described in the text, and mentioned in the top right corner of each plot. The solid black lines represent the DM relic density, while the black dotted curves indicate the DM equilibrium density. The black dashed vertical lines mark the scale factor corresponding to freeze-out temperature ($\rm T_{fo}$) and reheating temperature ($\rm T_{rh}$). The horizontal red dashed lines highlight the observed DM relic density.}
    \label{fig:relicDM}
\end{figure}
In Fig.~\ref{fig:relicDM}, we present the evolution of the DM relic density in terms of scale factor for two benchmark points consistent with the observed DM relic abundance. The evolution of the DM relic is obtained by solving the coupled BEQs Eqs.~\ref{eq:phieom},~\ref{eq:Reom}, and~\ref{eq:BEQDM}, along with using the relation $\mathcal{H}=\sqrt{(\rho^{}_{R}+\rho_{\Phi}^{})/M_{Pl}^{}}$. As discussed earlier, owing to the limitations on the attainable centre-of-mass energy in current and future colliders, our analysis is restricted to the bosonic reheating scenarios. Fig.~\ref{fig:relicDM} correspond to the $n=4$ case, with the reheating temperature fixed and identical for both subplots, while $m_{\phi}$ and $\Lambda$ are varied to obtain the correct DM relic density (see Eq.~\ref{eq:Omegarh}). For a fixed reheating temperature, an increase in the DM mass leads to a higher freeze-out temperature, thereby widening the gap between $T_{\rm rh}^{}$ and $T_{\rm fo}^{}$. This larger gap enhances the entropy dilution effect, reducing the DM relic abundance. To counterbalance this reduction and recover the observed DM relic density, the DM annihilation cross-section must be decreased, which can be achieved by increasing $\Lambda$ (see Eq.~\ref{eq:sigmavan}). Thus, due to the entropy-dilution effect, a smaller annihilation cross section (a larger new-physics scale) can yield the correct relic abundance. It is also worth mentioning that the bump in each plot of the figure is due to the change in the number of relativistic degrees of freedom during the QCD phase transition. The relic density evolution for $n=6$ (and higher) case is discussed in Appendix~\ref{app:dmn6}, where the DM never stays in equilibrium and hence can not be classified as a WIMP. For this reason, we restrict our analysis to the $n=4$ bosonic reheating scenario only.  
\begin{figure}[htb!]
	\centering
	\includegraphics[width=0.75\linewidth]{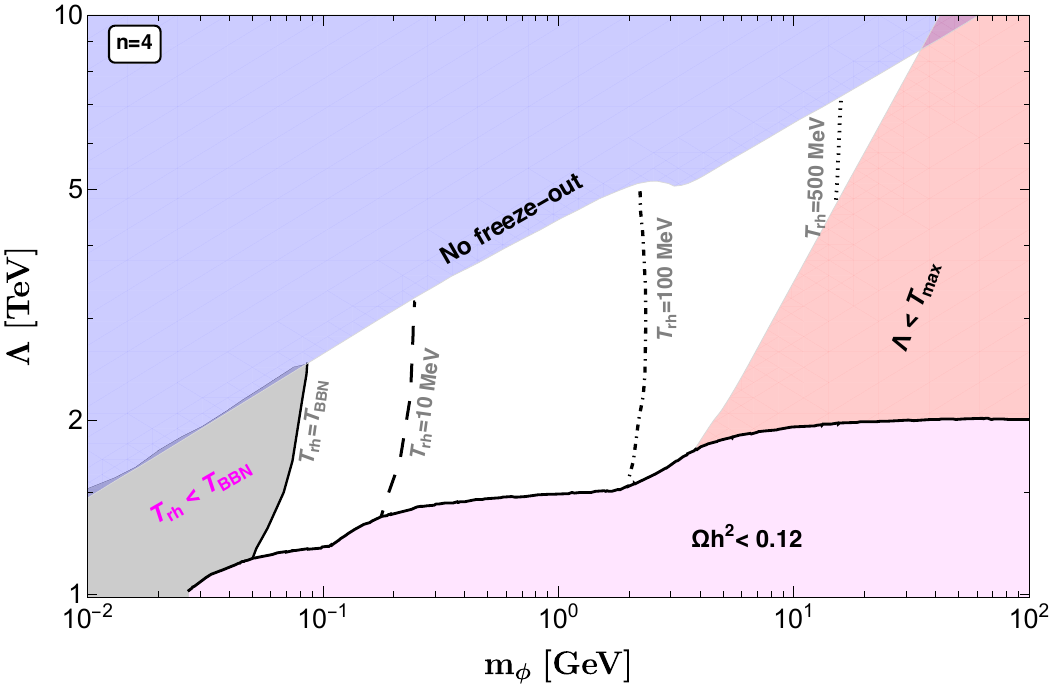}
	\caption{Viable parameter space for dimension-6 leptophilic interaction-driven (see Eq.~\ref{eq:EFToptour}) DM freeze-out during the reheating epoch (with $n=4$ bosonic reheating). Contours corresponding to the correct DM relic abundance are shown for $\rm T_{rh}=T_{BBN}$ (\textit{solid} line), $\rm T_{rh}=10$ MeV (\textit{dashed} line), $\rm T_{rh}=100$ MeV (\textit{dash-dotted} line) and $\rm T_{rh}=500$ MeV (dotted line). The thick \textit{black} line represents the standard scenario where DM freezes out during the radiation-dominated era.}
	\label{fig:fo-dur-rh}
\end{figure}

In Fig.~\ref{fig:fo-dur-rh}, we have shown the contours of reheating temperatures required to achieve the observed DM abundance, in the $\{m_{\phi}^{},~\Lambda\}$ plane for the $n=4$ bosonic reheating scenario. The thick black line represents the standard scenario, where the DM freezes out during the RD epoch, while the thin black lines correspond to $T_{\rm rh}^{}=T_{\rm BBN}^{}$, $10$ MeV, $100$ MeV, and $500$ MeV. The shaded regions in the plot indicate specific constraints: {\tt(i)} The purple-shaded region below the thick black line corresponds to scenarios with large DM annihilation cross-sections, where DM decouples too late from the thermal bath, resulting in an underabundant DM relic ($\Omega_{\rm DM}^{}h^2<0.12$). {\tt(ii)} To respect the BBN predictions, the reheat temperature must not go below the BBN temperature (excluded by the dark-shaded region). This condition creates a lower bound on the DM mass, $m_{\phi}^{}\gtrsim 10^{-1}$ GeV. {\tt(iii)} DM can not achieve chemical equilibrium with the visible sector if the thermal average annihilation cross-section is too small. This region of parameter space with higher values of $\Lambda$ is known as the ``No freeze-out'' zone (depicted by the blue shaded region), where the observed DM relic can still be achieved (see the lower row of Fig.~\ref{fig:relicDM}); however, in this case, the DM production mechanism is similar to the FIMP scenario. And {\tt(iv)}, the effective description of the DM-SM interaction must remain valid throughout the analysis, which requires satisfying the condition $\Lambda>T_{\rm max}^{}$. This constraint is illustrated by the red shaded region of the plot.
\section{Constraints on Model Parameters}
\label{sec:constraints}
In this section, we examine the existing experimental and observational bounds that constrain the parameter space of our DM model. These include terrestrial limits from direct and indirect detection experiments, astrophysical constraints from supernova observations, and collider bounds from previous accelerator data. Together, they serve to narrow down the viable regions in the $m_{\phi}-\Lambda$ parameter space and set the stage for evaluating the discovery prospects at future colliders.
\subsection{Direct Detection}
\label{sec:dd}
In the sub-GeV dark matter (DM) mass regime, the dominant contribution to atomic recoil arises from DM interactions with bound atomic electrons. The cross section for the elastic scattering process $e\phi \to e\phi$ is given by~\cite{Giagu:2019fmp}  
\begin{equation}
    \bar{\sigma}_{e\phi \to e\phi} \equiv \frac{\mu^{2}_{\phi e}}{16\pi m_{\phi}^{2}m^{2}_{e}}\overline{|\mathcal{M}_{e\phi \to e\phi}({\bf q})|^2}\bigg|_{|{\bf q}|^{2}=\alpha^2m_{e}^{2}}\,.
\end{equation}
Here, $\mu_{\phi e}$ is the reduced mass of the DM-electron system. In the limit $m_{\phi}^{}\gg m_{e}$, this DM electron recoil cross-section is expressed as  
\begin{equation}
    \bar{\sigma}_{e\phi \to e\phi}\approx\frac{57}{16\pi(m_{e}+m_{\phi})^{2}}\frac{m_{e}^{2}~m_{Z}^{2}}{\Lambda^{4}}\,,
\end{equation}
where $m_{Z}^{}$ is the mass of the Z boson. Direct detection experiments such as XENON1T~\cite{XENON:2019gfn}, PandaX-4T~\cite{PandaX:2022xqx}, and others have placed upper bounds on the electron-DM scattering cross section. Among these, PandaX-4T provides the strongest constraints. However, these bounds exclude only a narrow region of the $m_\phi - \Lambda$ parameter space near $m_\phi \approx 100\,\mathrm{MeV}$. Most of the viable region corresponds to under-abundant relic density; therefore, the direct detection limits do not significantly impact our analysis, therefore not visible in Fig.~\ref{fig:fo-dur-rh}. 
\subsection{Indirect Detection}
\label{sec:id}
The annihilation cross section of dark matter into a charged lepton pair $\ell^+ \ell^-$ is given in Eq. \ref{ID_eq1}. For the $e^+e^-$ final state, the electron mass can be neglected compared to the CM energy ($m_e \ll \sqrt{s}$), simplifying the equation as follows,
\begin{equation}
	\sigma = \frac{v_{\rm EW}^{2}}{4\pi \Lambda^{4}} \left(1 - \frac{4 m^{2}_{\phi}}{s}\right)^{-\frac{1}{2}}~.
    \label{ID_eq2}
\end{equation}
In the CM frame, the momenta of the two annihilating dark matter particles are equal, $|\mathbf{p}_1| = |\mathbf{p}_2| = p = \frac{m_\phi v}{2}$, where $v$ is the relative velocity of the dark matter particles in the CM frame. Their energies are also equal, $E_1 = E_2 = E$, leading to $s = 4E^2$. In the non-relativistic limit, where $\frac{p^2}{m_\phi^2} \ll 1$, the energy simplifies to
\begin{equation}
E =\sqrt{p^2 + m_\phi^2} \simeq m_\phi \left(1 + \frac{p^2}{2m_\phi^2}\right) \simeq m_\phi \left(1 + \frac{v^2}{8}\right)~.
    \label{ID_eq3}
\end{equation}
Thus,
\begin{equation}
s = 4 E^2=    4 m_\phi^2 \left(1 + \frac{v^2}{4}+\mathcal{O}(v^4) \right)~.
    \label{ID_eq4}
\end{equation}
Substituting Eq. \ref{ID_eq4} into Eq. \ref{ID_eq2} and expanding in the limit where the relative velocity $v$ is much less than the speed of light (in natural units, $\hbar=c=1$), we obtain the following expression for the annihilation cross section
\begin{equation}
	\langle\sigma v\rangle = \frac{v_{\rm EW}^{2}}{2\pi \Lambda^{4}} \left(1 + \frac{v^2}{8}+\mathcal{O}(v^4)\right)\,,
    \label{ID_eq5}
\end{equation}
where the relative molar velocity of the DM particles is denoted by $v$, this is around $v\sim 10^{-3}$ (in natural units) inside the Milky Way halo. We include the DM indirect detection (ID) constraint from the CMB observation (arising from DM annihilation to charged particles)~\cite{Planck:2018vyg, Madhavacheril:2013cna, Slatyer:2015jla}, this is the most stringent bound for DM mass below $5$ GeV. We also use the limits on the DM annihilation cross-section from Alpha Magnetic Spectrometer-$02$ (AMS-$02$)~\cite{AMS:2013fma, Boudaud:2014dta}. Currently, this observation provides the most precise data for charge cosmic ray measurements, especially in the $10-1000$ GeV DM mass range. The exclusion contours from the CMB and (AMS-$02$) data on the dimension-six leptophilic operator, in the $m_\phi$–$\Lambda$ parameter space, are shown as shaded grey regions in Fig.~\ref{fig:fo-dur-rh-const}.
\subsection{Supernova Bounds}
\label{sec:sn}
Significant production of leptophilic DM can occur inside the supernova core through electron-positron annihilation, provided the dark matter mass is of the same order as the core temperature, $T_{c}\sim\mathcal{O}(30)$ MeV. If the dark matter's mean free path is comparable to the supernova core radius, $R_{c}\sim\mathcal{O}(10)$ km, the DM particles can freely escape, thereby enhancing the supernova cooling rate. Observations of $\rm SN1987A$ impose stringent limits on such additional cooling rates~\cite{Kamiokande-II:1987idp, PhysRevLett.58.1494}. 
To quantify this effect, we adopt the Raffelt criterion~\cite{Raffelt:1996wa}, which constrains the total energy loss rate per unit mass of the supernova via the following condition on emissivity
\begin{equation}
    \dot\epsilon<10^{19} \rm erg~g^{-1}~s^{-1}\,,
\end{equation}
and follow the methodology outlined in ~\cite{Dreiner:2003wh, Guha:2018mli, Magill:2018jla} to translate this bound into constraints on the DM mass and NP scale $\Lambda$. 
In our scenario, the emissivity quantifies the total energy emitted per unit time and unit volume from the supernova due to the process $e^{+}(p_{1})+e^{-}(p_{2})\rightarrow\phi(k_{1})+\phi(k_{2})$ and can be expressed as~\cite{Dreiner:2003wh} 
\begin{equation}
    \dot{\epsilon}(m_{\phi}^{},~T_{c}^{},~\frac{\mu}{T_{c}^{}})=\frac{1}{\rho_{\rm SN}^{}}\int \frac{d^{3}p_{1}d^{3}p_{2}}{(2\pi)^6}(E_{1}+E_{2})f_{1}f_{2}v_{M}^{}\sigma\,,
\end{equation}
where $\mu$ is the chemical potential of the electron, $\rho_{\rm SN}^{}=3\times10^{14}\text{g}~\text{cm}^{-3}$ is the supernova core density, and $v_{M}^{}$ is the Møller relative velocity. The annihilation cross-section, $\sigma$, is identical to that in Eq.~\ref{ID_eq1}, except that $m_{\ell}^{}$ is replaced by $m_{e}^{}$. $f_1$ and $f_2$ denote the Fermi-Dirac distributions of the incoming positrons and electrons, evaluated at the supernova core temperature $T_c$.

\begin{figure}[htb!]
	\centering
	\includegraphics[width=0.85\linewidth]{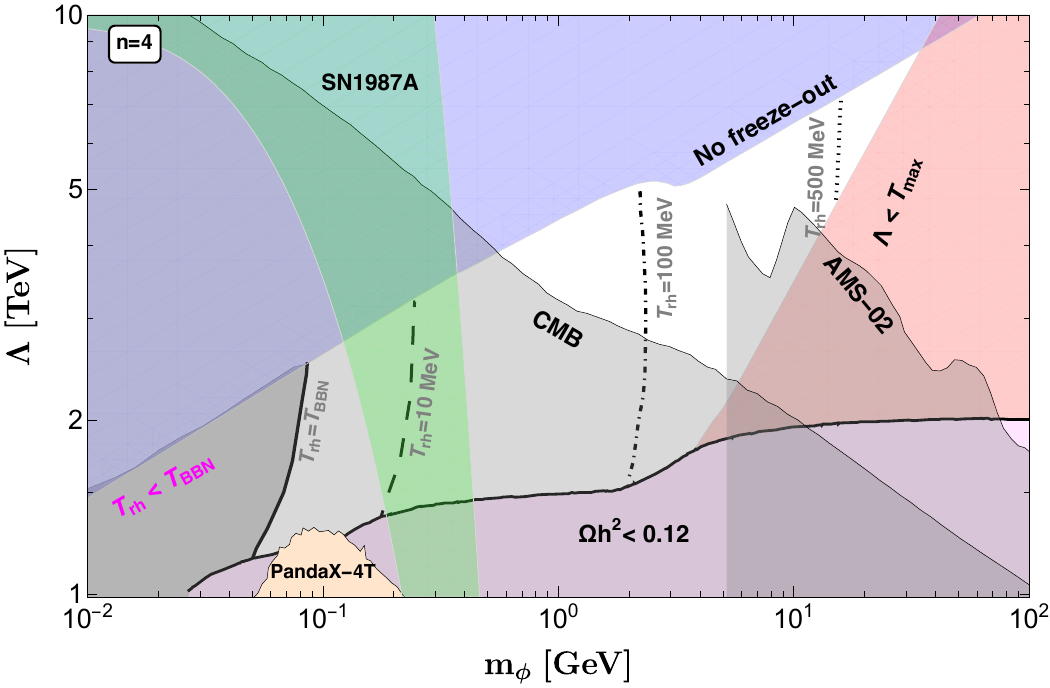}
	\caption{Modified parameter space with various observational bounds for leptophilic DM which freezes out during the reheating era (with $n=4$ bosonic reheating scenario).}
	\label{fig:fo-dur-rh-const}
\end{figure}

The Raffelt criterion is relevant when the DM particle that is produced inside the supernova can freely escape from its core, thereby carrying away some portion of its energy. In scenarios with stronger interactions between DM and electron, the corresponding mean free path of the DM ($\lambda_{\phi}$) can be substantially reduced. For DM scattering with supernova electrons, the mean free path is given by
\begin{equation}
    \lambda_{\phi}^{}=\frac{1}{n_{e}\sigma_{e\phi \to e\phi}}\,,
\end{equation}
where $n_{e}$ denotes the electron number density inside the core of the supernova and $\sigma_{e\phi \to e\phi}$ represents the DM-electron scattering cross-section. The condition for DM to free stream out from the supernova core is evaluated using the optical depth criterion:
\begin{equation}
    \int_{0.9R_{c}}^{R_{c}}\frac{dr}{\lambda_{\phi}}\lesssim\frac{2}{3}\,.
\end{equation}
It is important to note that when the DM mass exceeds the average temperature of the supernova core, the inverse of the mean free path must be scaled by the Boltzmann suppression factor $e^{-E_{\phi}/T}$ to account for the reduced DM number density at larger mass. In Fig.~\ref{fig:fo-dur-rh-const}, the corresponding exclusion contour from supernova cooling data is shown by shaded green regions. 

\subsection{Limits from LEP}
\label{sec:lep}
\begin{figure}[htb!]
    \centering
    \includegraphics[width=0.465\linewidth]{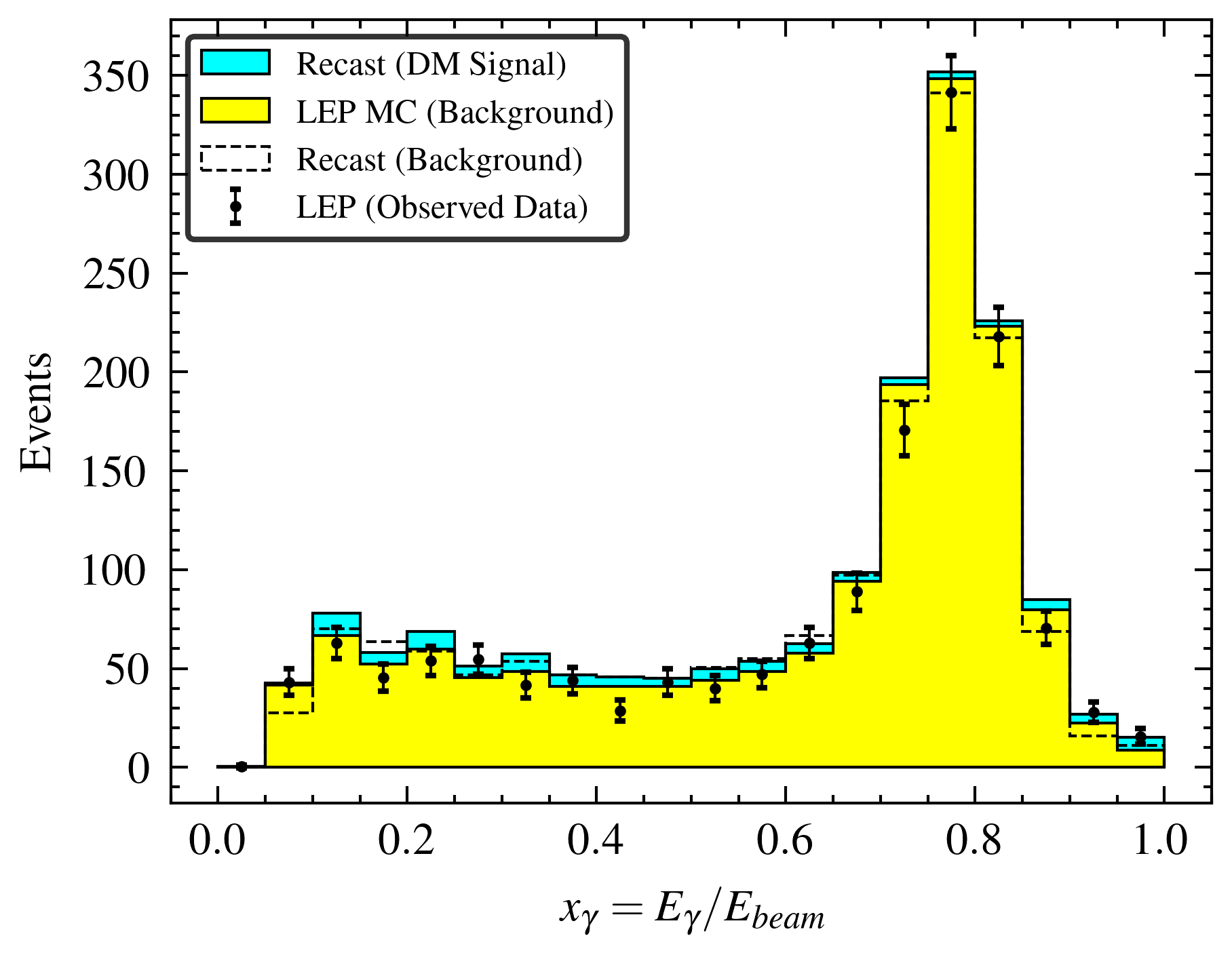}
    \includegraphics[width=0.475\linewidth]{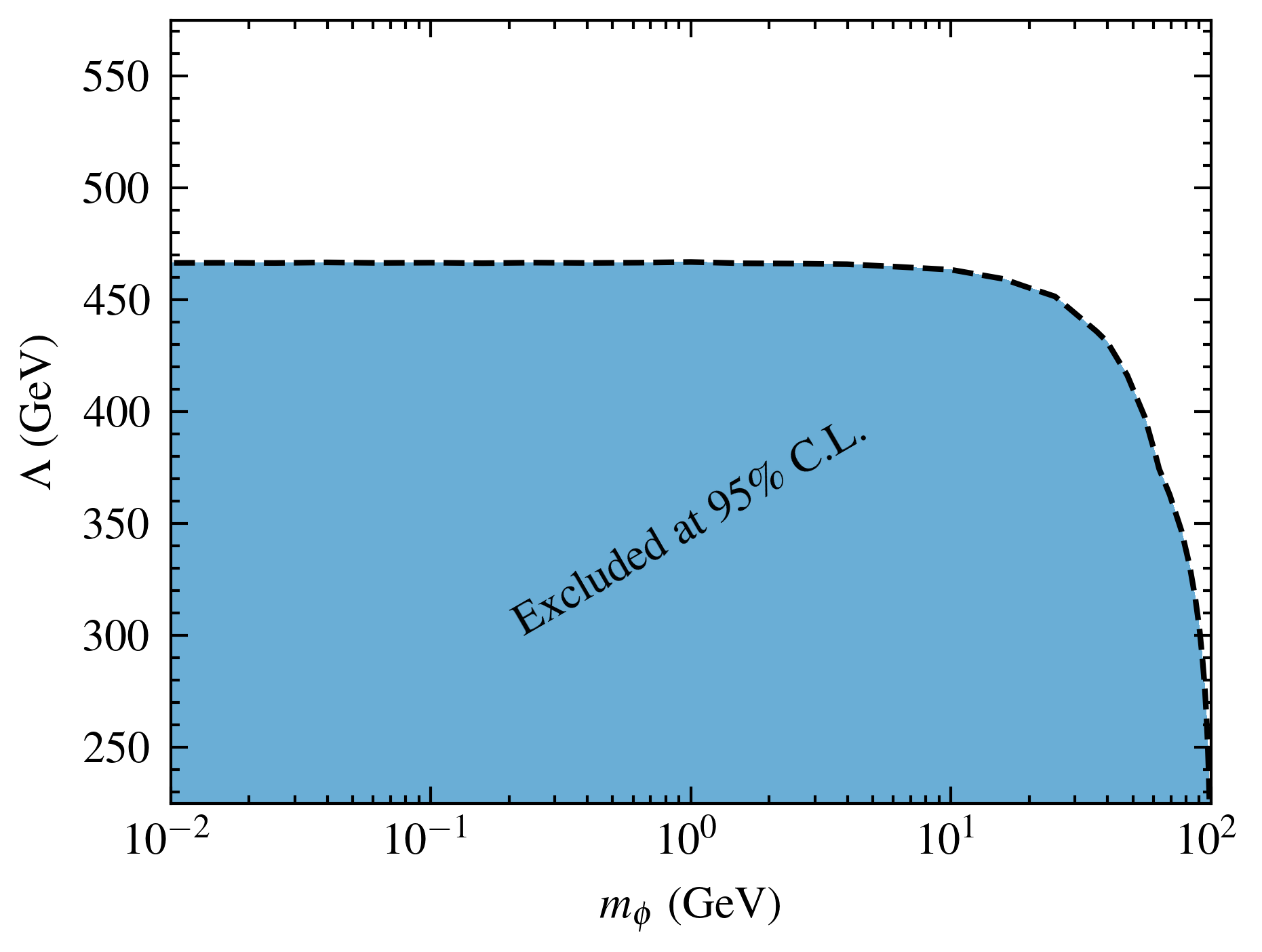}
    \caption{{\it Left:} Binned $x_{\gamma}$ distribution for LEP studies corresponding to $\mathfrak{L}_{\rm int}$ = 650 pb$^{-1}$. The DM signal corresponds to $\phi \phi \gamma$ production for the benchmark $m_{\phi} = 5$ GeV, $\Lambda = 2.5$ TeV. {\it Right:} The shaded region is excluded at 95\% C.L. (bordered: \textit{black, dashed}) from the LEP recast study.}
    \label{fig:lep}
\end{figure}

The Large Electron–Positron Collider (LEP) conducted extensive studies of mono-photon events with missing energy~\cite{DELPHI:2003dlq, DELPHI:2008uka}. However, most of them were aimed towards precision measurements within the SM or other exotic searches. One way to extract information from these existing studies is to recast them using identical set of resolutions and efficiencies as used in the original studies and feed Monte Carlo (MC) generated events through the recasting framework. Such study has been done concerning fermionic leptophilic operators, for example, see Ref.~\cite{Fox:2011fx}. Here we provide an exclusion bound on the $m_{\phi}-\Lambda$ plane based on existing LEP data from mono-$\gamma$ studies. The mono-$\gamma$ signal for our operator is shown in Fig.~\ref{fig:dm1} (\textit{left}). The dominant SM background comes the $\nu \overline{\nu} \gamma$ process. It should be noted that in the low DM mass regime, the cross section and photon energy distributions are nearly independent of the DM mass. Hence, this limits cater to the entire low mass regime. LEP studies were done over a range of CM energies $\sqrt{s}=[180-209]$ GeV. The observed events, however, were reconstructed and presented in bins of $x_{\gamma}$, defined as $E_{\gamma} / E_{\rm beam}$, thus independent of CM energy of the collisions. For our analysis, we generate events at a reference CM mass energy of $\sqrt{s} = 200$ GeV, which, following the conclusion drawn in~\cite{Fox:2011fx}, does not invalidate the LEP data. The DELPHI detector at LEP had three angular regions (HPC, FEMC and STIC), and each region had a different set of trigger and reconstruction/identification efficiencies. The recasting of each of these regions, based on~\cite{DELPHI:2003dlq, DELPHI:2008uka}, are detailed in Tab.~\ref{tab:lep}. There is an additional photon identification efficiency of 90\% valid for all regions. For STIC, since, the information provided by~\cite{DELPHI:2003dlq} is incomplete, we assume the overall efficiency to be 48\% as done by \cite{Fox:2011fx}. In order to validate our framework, we plot the $x_{\gamma}$ distribution of our generated MC events for the SM backgrounds on top of the same from DELPHI MC studies. They are found to be in perfect agreement as shown in Fig.~\ref{fig:lep}. The observed data as well as the DM signal corresponding to the benchmark $m_{\phi} = 5$ GeV, $\Lambda = 2.5$ TeV.
\begin{table}[htb!]
\centering
\renewcommand{\arraystretch}{1}{\small
\begin{tabular}{|c|c|c|c|}
\hline
Detector  & Trigger & Reconstruction & Energy \\
Regions & Efficiency & Efficiency & Smearing \\ \hline
& $x_{\gamma} \in [0.06, 0.30]$ &  $x_{\gamma} \in [0.06, 0.80]$ & Gaussian ($\sigma_{E}/E_{\gamma}$): \\
HPC & (45.75 + 1.042 $E_{\gamma}$)\% &  (38 + 0.5 $E_{\gamma}$)\% & $ 0.043 \oplus 0.32/ \sqrt{E_{\gamma}} $ \\
$\theta_{\gamma} \in [90^{\circ}, 45^{\circ}]$  & $\cup$ & $\cup$ & + \\
$x_{\gamma} \in [0.06, 1.10]$  & $x_{\gamma} \in [0.30, 1.10]$ &$x_{\gamma} \in [0.80, 1.10]$  & Lorentzian ($\Gamma$): \\
& (74 + 0.1 $E_{\gamma}$)\% & 78\% & 0.04 $E_{\gamma}$ \\
\hline
& $x_{\gamma} \in [0.10, 0.15]$ &  & Gaussian ($\sigma_{E}/E_{\gamma}$): \\
FEMC & (79 + 1.4 $E_{\gamma}$)\% &  & $ 0.03 \oplus 0.12/ \sqrt{E_{\gamma}} \oplus 0.11/ E_{\gamma} $  \\
$\theta_{\gamma} \in [32^{\circ}, 12^{\circ}]$  & $\cup$ & $0.89(55 + 0.2 E_{\gamma})\%$ & + \\
$x_{\gamma} \in [0.10, 0.90]$  & $x_{\gamma} \in [0.15, 0.90]$ &  & Lorentzian ($\Gamma$): \\
& 100\% &  & 0.02 $E_{\gamma}$ \\
\hline
& \multicolumn{2}{c|}{} & Gaussian ($\sigma_{E}/E_{\gamma}$): \\
STIC & \multicolumn{2}{c|}{} & $0.0152 \oplus 0.135/ \sqrt{E_{\gamma}}$ \\
$\theta_{\gamma} \in [8^{\circ}, 3.8^{\circ}]$  & \multicolumn{2}{c|}{48\%} & + \\
$x_{\gamma} \in [0.30, 0.90]$  & \multicolumn{2}{c|}{} & Lorentzian ($\Gamma$):  \\
& \multicolumn{2}{c|}{} & 0.02 $E_{\gamma}$ \\
\hline
\end{tabular}}
\caption{Details of resolution and efficiencies for the LEP recast study. The $\theta_{\gamma}$ ranges are shown for one half of the detector only, but the same efficiencies apply for the other half as well. Additionally we implement angular cuts: $\theta_{\gamma} > (28 - 80 x_{\gamma})^{\circ}$ and $\theta_{\gamma} > (9.2 - 9 x_{\gamma})^{\circ}$ for FEMC and STIC regions respectively. For further details, see Ref.~\cite{Fox:2011fx}.}
\label{tab:lep}
\end{table}

To obtain an exclusion bound on $\Lambda$, we perform a $\Delta \chi^{2}$ test for the binned $x_{\gamma}$ distribution. We exclude the first bin and consider other 19 bins for the analysis. The $\Delta \chi^{2}$ for this case is defined as:
\begin{equation}
\Delta \chi^{2} (\Lambda) = \sum^{19}_{i = 1} \left(\frac{N_{\rm obs} - (S (\Lambda) + B)}{\sqrt{S(\Lambda) + B}}\right)^{2},
\end{equation}
where, $N_{\rm obs}$ is the number of events observed in each bin, $S$ and $B$ are number of signal and background events post detector efficiencies. The degrees of freedom (dof) for the binned $\Delta \chi^{2}$ analysis is $N - M = 19 - 2 = 17$, where, $N$ is the number of bins and $M$ is the number of model parameters. For 17 dofs, the $\Delta \chi^{2}$ value corresponding to 95\% C.L. is 27.587. Fig.~\ref{fig:lep} shows, the 95\% C.L. exclusion contour in the $m_{\phi}-\Lambda$ plane.

\section{DM Searches at Lepton Collider}
\label{sec:collider}
\begin{figure}[htb]
	\centering
	\begin{tikzpicture}[baseline={(current bounding box.center)},style={scale=1, transform shape}]
	\begin{feynman}
		\vertex[square dot](a){};
		\vertex[above left =0.625cm and 1cm of a] (a1);
            \vertex[above right =0.625cm and 1cm of a1] (a4){$\gamma/Z$};
            \vertex[above left =0.45cm and 0.75cm of a1] (a3){$e^{-}$};
		\vertex[below left =1.25cm and 2cm of a] (a2){$e^{+}$};
		\vertex[above right = 1.25cm and 2cm of a] (b1) {$\phi$};
		\vertex[below right = 1.25cm and 2cm of a] (b2){$\phi$};
		\diagram*{
			(a) -- [thick, fermion, arrow size=1pt] (a2),
			(a1) -- [thick, fermion, arrow size=1pt] (a),
                (a3) -- [thick, fermion, arrow size=1pt] (a1),
                (a1) -- [thick, boson, arrow size=1pt] (a4),
			(a) -- [thick, scalar, arrow size=0.7pt] (b1),
			(b2) -- [thick, scalar, arrow size=0.7pt] (a)
		};
	\end{feynman}
	\end{tikzpicture}
	\hspace{1cm}
	\begin{tikzpicture}[baseline={(current bounding box.center)},style={scale=1, transform shape}]
		\begin{feynman}
			\vertex[square dot](a){};
			\vertex[above left =1.25cm and 2cm of a] (a1){$e^{-}$};
			\vertex[below left =1.25cm and 2cm of a] (a2){$e^{+}$};
			\vertex[above right = 1.25cm and 2cm of a] (b1) {$\phi$};
			\vertex[below right = 1.25cm and 2cm of a] (b2){$\phi$};
			\vertex[right = 2cm of a] (b3){$h$};
			\diagram*{
				(a) -- [thick, fermion, arrow size=1pt] (a2),
				(a1) -- [thick, fermion, arrow size=1pt] (a),
				(a) -- [thick, scalar, arrow size=0.7pt] (b1),
				(a) -- [thick, scalar] (b3),
				(b2) -- [thick, scalar, arrow size=0.7pt] (a)
			};
		\end{feynman}
	\end{tikzpicture}
	\caption{Feynman diagrams corresponding to mono-$\gamma/Z$ (\textit{left}) and mono-$h$ signal at $e^{+}e^{-}$ colliders (\textit{right}). The square dot corresponds to the EFT vertex, for example, one considered in Eq.~\ref{eq:EFToptour}.}
    \label{fig:dm1}
\end{figure}
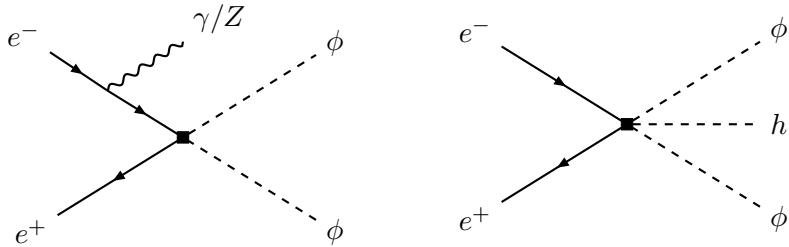

Lepton colliders offer a clean experimental environment for probing DM and dark sector particles, with significant advantages over hadron colliders. In contrast to hadron machines, where large QCD backgrounds, proton remnants, and pile-up obscure missing energy signals, lepton colliders provide well-defined initial states, negligible QCD activity, and virtually no pile-up. This enables precise reconstruction of missing energy (inaccessible at hadron colliders) and missing transverse momentum, crucial observables in DM searches. The clean conditions also allow for model-independent mono-$X$ searches ($X = \gamma, Z, W, h, j$), where DM recoils against a visible object. These channels greatly enhance sensitivity to weakly interacting dark sector particles, making lepton colliders a powerful complement to hadron colliders.

In this section, we discuss collider probes of the leptophilic DM scenario under consideration, focusing on searches at the ILC. Several studies have explored lepton collider signatures of DM via mono-$\gamma/Z$ channels~\cite{Fox:2011fx,Yu:2013aca,Essig:2013vha,Yu:2014ula,Freitas:2014jla,Dutta:2017ljq,Horigome:2021qof,Barman:2021hhg,Kundu:2021cmo,Bhattacharya:2022qck,Ge:2023wye,Ma:2022cto}. These searches rely on $\gamma/Z$ radiation from the initial state electron-positron pair. However, a major challenge in this context is distinguishing such signals from the irreducible SM neutrino background, as demonstrated in~\cite{Barman:2024nhr}. As a result, feeble DM signals are often highly suppressed, limiting the reach of such searches to scenarios with large signal excesses. In our case, due to the Yukawa-like structure of the effective operator, the DM signal is not limited to mono-$\gamma/Z$ channels. The operator also induces a mono-$h$ signal through a contact interaction, as illustrated in Fig.~\ref{fig:dm1} (\textit{right}). This channel exhibits distinctive features compared to SM background processes, which we will detail later. Importantly, the contact nature of the interaction leads to an enhancement of the signal cross section with increasing center-of-mass energy. Hence, we perform our analysis at the maximum energy reach of the ILC, i.e. $\sqrt{s} = 1$~TeV, assuming an integrated luminosity of $\mathfrak{L}_{\rm int} = 8$~ab$^{-1}$. Mono-Higgs searches at the ILC offer a clean and distinctive signature for probing DM via recoil against a visible Higgs boson, enabling precise reconstruction of the missing mass and energy. The well-defined initial state and beam polarization further enhance the sensitivity to DM–Higgs interactions while suppressing backgrounds.

For the signal, we reconstruct the mono-$h$ channel through the dominant decay mode of the Higgs boson, namely, $h \to b\overline{b}$. The dominant SM background arises from the processes $\nu \overline{\nu}h(b \overline{b})$ and $\nu \overline{\nu} + \text{jets}$. In the latter case, we include all jet flavors—bottom, charm, and light jets, due to the possibility of mis-tagging, which can mimic the $b\overline{b}$ final state. The representative Feynman diagrams for the SM background are shown in Appendix~\ref{sec:SMBG} (Fig.~\ref{fig:BG}).
Another feature of lepton colliders like the ILC is the ability to polarize the initial state $e^+e^-$ beams. Beam polarization provides a powerful handle to enhance signal sensitivity and suppress background processes. For leptophilic DM scenarios, where interactions involve specific chiral structures, choosing suitable polarization configurations can significantly boost the signal cross section while reducing contributions from SM backgrounds, particularly those involving neutrinos. This leads to improved discrimination between signal and background, thereby extending the reach of DM searches. Operators with Yukawa-like structure combine fermions of different chiralities, hence specific polarization choices can significantly enhance signal contribution while diminishing the backgrounds. The ILC run at $\sqrt{s} = 1$ TeV, projects a polarization of upto $\pm 80\%$ for the electron beams, while at least upto $\pm 20\%$ polarization is possible for the positron beam~\cite{Behnke:2013xla}.

The model implementation of the DMEFT operator is carried out using \texttt{FeynRules}~\cite{Alloul:2013bka}, and the corresponding Universal FeynRules Output (UFO)~\cite{Degrande:2011ua} is imported into \textsc{MadGraph5\_aMC@NLO}~\cite{Alwall:2014hca,Frederix:2018nkq}. Production cross sections for both the signal and relevant SM backgrounds, for various beam polarization configurations, are presented in Tab.~\ref{tab:prodxs}. The DM signal corresponds to the benchmark: $m_{\phi} = 5$ GeV and $\Lambda = 2.5$ TeV. These cross sections are computed at the parton level using \textsc{MadGraph5\_aMC@NLO}, without applying any additional selection criteria. The Higgs decay to a bottom quark pair is incorporated by multiplying the total production cross section with the corresponding branching ratio. Among the polarization configurations considered, $\left(P_{e^+}, P_{e^-}\right) = \left(+20\%, +80\%\right)$ emerges as the most favorable choice, as it enhances the signal rate while significantly suppressing the SM background. The $W$ boson in the Standard Model couples only to left-handed fermions and right-handed antifermions. Under the beam polarization $(P_{e^+}, P_{e^-}) = (+20\%, +80\%)$, the electron beam is predominantly right-handed, leading to a strong suppression of the left-handed electron component necessary for the $t$-channel $W$ exchange processes. Consequently, the dominant contributions to SM background processes such as $e^+ e^- \to \nu_e \bar{\nu}_e h$ and $e^+ e^- \to \nu \bar{\nu} jj$ are significantly reduced. It is obvious that with higher degrees of polarisation of the positrons in particular, the significance is going to get better. We perform our detailed analysis using this polarization setup, while also including results for the unpolarized case as a reference.

\begin{table}[htb!]
    \centering
    \renewcommand{\arraystretch}{1.0}{
    \begin{tabular}{|>{\centering\arraybackslash}p{3cm}|
                     >{\centering\arraybackslash}p{2cm}|
                     >{\centering\arraybackslash}p{2cm}|
                     >{\centering\arraybackslash}p{2cm}|}
    \hline 
    Polarization  & \multicolumn{3}{c|}{Production cross section (fb)} \\ \cline{2-4}
    $\left(P_{e^+}, P_{e^-}\right)$ & $\phi \phi h(b\overline{b})$ & $\nu \overline{\nu} h(b\overline{b})$ & $\nu \overline{\nu} + {\rm jets}$ \\ \hline
    Unpolarized & 0.1590 & 120.8 & 512.0 \\
    $\left(+20\%, +80\%\right)$ & 0.1841 & 29.69 & 139.5 \\
    $\left(+20\%, -80\%\right)$ & 0.1334 & 259.3 & 1081 \\
    $\left(-20\%, +80\%\right)$ & 0.1334 & 20.61 & 109.0 \\
    $\left(-20\%, -80\%\right)$ & 0.1841 & 172.7 & 723.7 \\ \hline
    \end{tabular}}
    \caption{Cross section of DM production in associated with mono-$h$, along with SM backgrounds, for different polarization combinations, at the ILC $\sqrt{s} = 1$ TeV run. The DM signal corresponds to the benchmark: $m_{\phi} = 5$ GeV and $\Lambda = 2.5$ TeV.}
    \label{tab:prodxs}
\end{table}

\subsection{Cut-Based Analysis}
\label{sec:cuts}
\begin{figure}[htb!]
    \centering
    \includegraphics[width=0.475\linewidth]{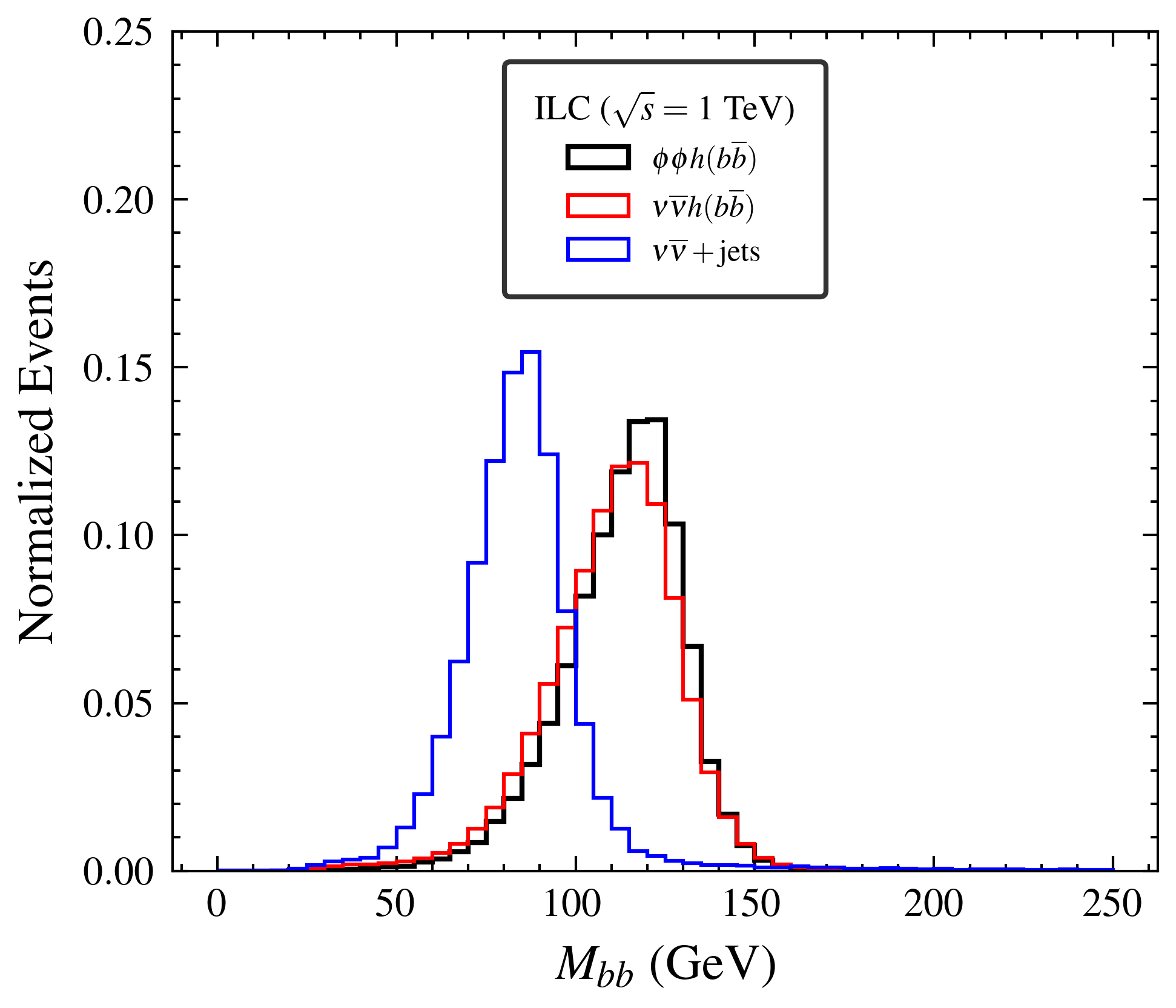}
    \includegraphics[width=0.475\linewidth]{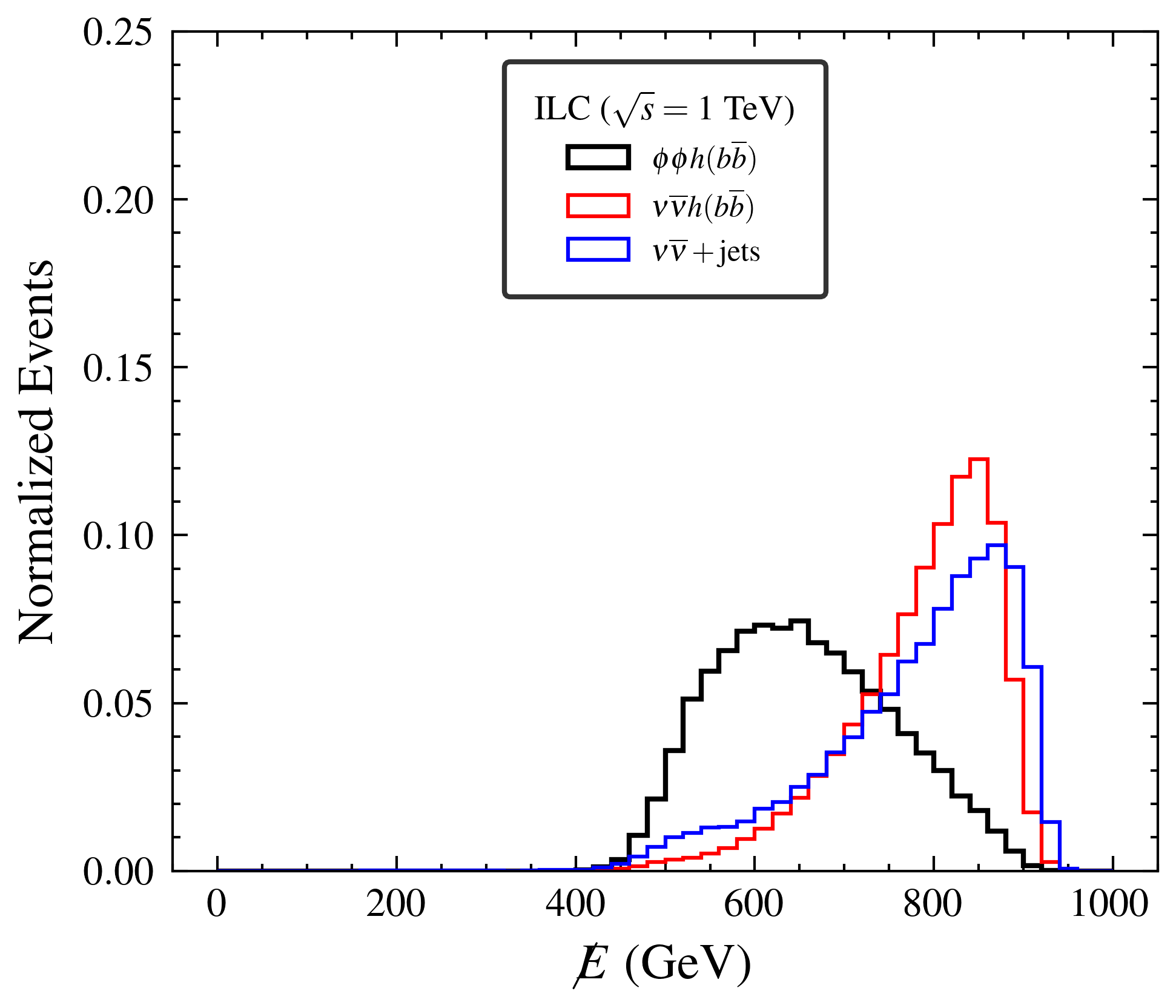}
    \includegraphics[width=0.475\linewidth]{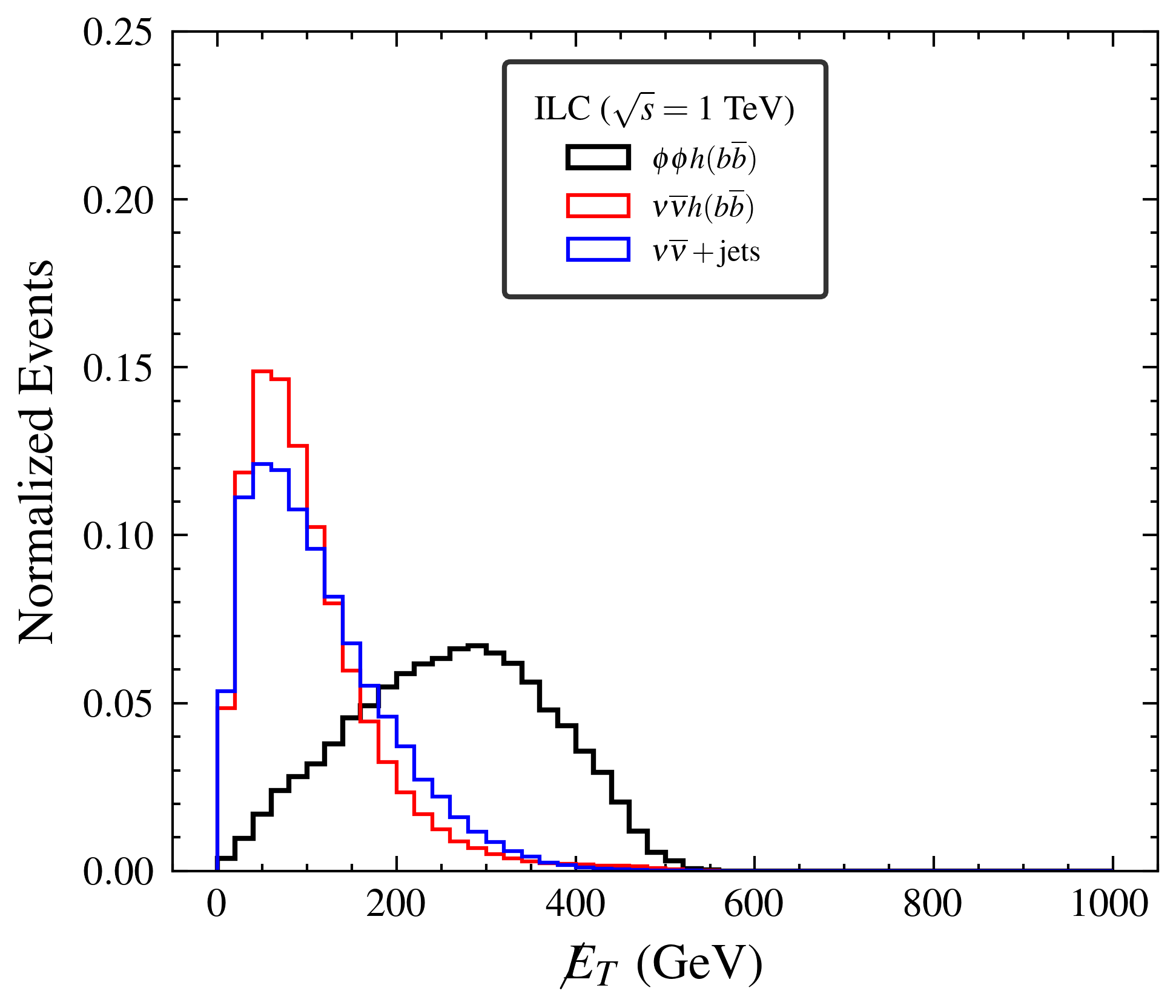}
    \includegraphics[width=0.475\linewidth]{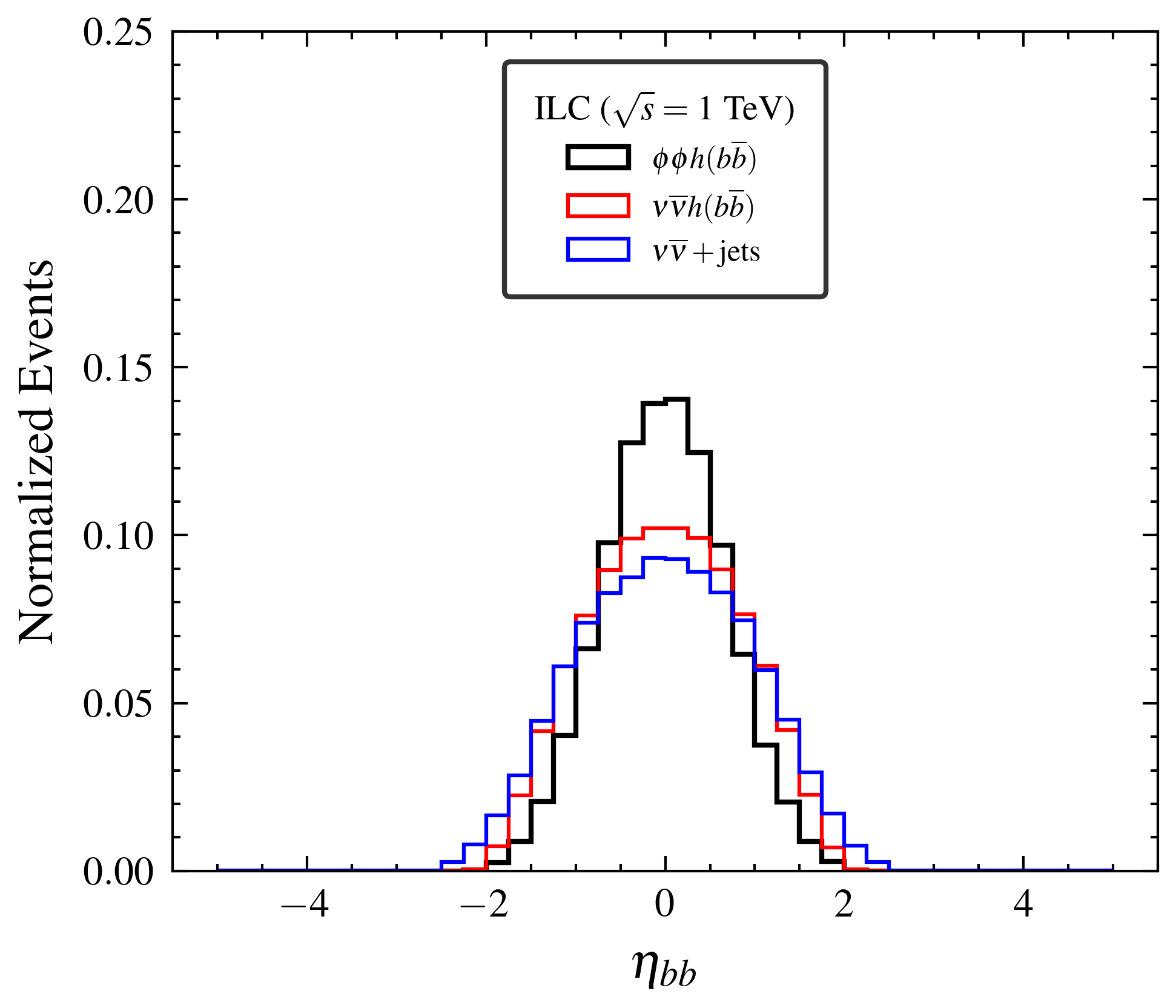}
    \caption{Normalized event kinematic distributions for signal and background processes. \textit{Top Left}: Invariant mass of $b$-jet pair ($M_{bb}$), \textit{Top Right}: Missing energy ($\slashed{E}$), \textit{Bottom Left}: Missing transverse momentum ($\slashed{E}_{T}$), \textit{Bottom Right}: pseudorapidity of $b$-jet pair system ($\eta_{bb}$).}
    \label{fig:dist}
\end{figure}

Monte Carlo events for both signal and background processes are generated using \textsc{MadGraph5\_aMC@NLO}. The generated parton-level events are passed to \texttt{Pythia8}~\cite{Bierlich:2022pfr} for parton showering, followed by detector simulation using \texttt{Delphes3}~\cite{deFavereau:2013fsa}. The detector resolution, tagging efficiencies, and other relevant parameters are taken from the default Delphes card provided with \textsc{MadGraph5\_aMC@NLO}. The basic selection criteria are as follows: events must contain exactly two jets, both of which are required to be $b$-tagged. Jets are reconstructed using the anti-$k_T$~\cite{Cacciari:2008gp} algorithm with a radius parameter of $R = 0.4$, requiring a minimum transverse momentum of 20 GeV. Events containing isolated leptons or photons are vetoed. Several kinematic variables relevant for discriminating signal from background are plotted in Fig.~\ref{fig:dist}, and their definitions are provided below:
\begin{itemize}
    \item {Invariant mass of the $b$-jet pair:}
    \begin{equation}
        M_{bb} = \sqrt{\left(p_{b_{1}} + p_{b_{2}}\right)^{2}},
    \end{equation}
    where $p_{b_{1}}$ and $p_{b_{2}}$ denote the four-momenta of the leading and sub-leading $b$-jets, respectively.

    \item {Missing transverse momentum:}
    \begin{equation}
        \slashed{E}_{T} = \sqrt{\left(\sum_{\rm visible} p_{x}\right)^{2} + \left(\sum_{\rm visible} p_{y}\right)^{2}},
    \end{equation}
    where $p_{x}$ and $p_{y}$ are the $x$- and $y$-components of the momenta of all visible final-state particles.

    \item {Missing energy:}
    \begin{equation}
        \slashed{E} = \sqrt{s} - \sum_{\rm visible} E,
    \end{equation}
    where $E$ is the energy of each visible final-state particle, and $\sqrt{s}$ is the center-of-mass energy of the collision.

    \item {Pseudorapidity of the $b$-jet pair system:}
    \begin{equation}
        \eta_{bb} = - \ln{\left[\tan{\left(\frac{\theta_{bb}}{2}\right)}\right]},
    \end{equation}
    where $\theta_{bb}$ is the polar angle of the reconstructed $b$-jet pair system, calculated from the transverse momenta of the individual $b$-jets.
\end{itemize}
In addition to the basic selection and detector-level cuts, we apply the following sequential kinematic cuts:
\begin{equation}
\begin{split}
    \text{Cut 1:}&\quad |M_{bb} - M_{h}| < 25~\text{GeV}\,,\\
    \text{Cut 2:}&\quad \slashed{E}_{T} > 265~\text{GeV}\,,\\
    \text{Cut 3:}&\quad \slashed{E} < 670~\text{GeV}\,.
\end{split}
\end{equation}
Selecting an invariant mass window around the Higgs mass serves as an effective discriminator between Higgs-mediated and non-Higgs background processes. In particular, the $\nu \overline{\nu} + \text{jets}$ background is significantly reduced by Cut~1. Further, the signal and background events exhibit distinct distributions in both missing transverse momentum and missing energy, see Fig.~\ref{fig:dist}. Specifically, the signal tends to peak at higher $\slashed{E}_{T}$ and lower $\slashed{E}$ compared to the SM backgrounds. Therefore, applying Cuts~2 and~3 sequentially leads to substantial suppression of SM background events. The overall cut efficiency is summarized in Tab.~\ref{tab:cutflow}, for unpolarized and polarized setups.
\begin{table}[htb!]
    \centering
    \renewcommand{\arraystretch}{1.0}{
    \begin{tabular}{|>{\centering\arraybackslash}p{2.5cm}|
                     >{\centering\arraybackslash}p{2cm}|
                     >{\centering\arraybackslash}p{3cm}|
                     >{\centering\arraybackslash}p{2cm}|
                     >{\centering\arraybackslash}p{3cm}|}
    \hline 
    \multirow{3}*{Cuts} & \multicolumn{4}{c|}{Number of events} \\ \cline{2-5}
    & \multicolumn{2}{c|}{Unpolarized} & \multicolumn{2}{c|}{$\left(P_{e^+}, P_{e^-}= +20\%, +80\%\right)$} \\ \cline{2-5}
    & Signal ($S$) & Background ($B$) & Signal ($S$) & Background ($B$) \\ \hline
    \multirow{2}*{Basic cuts} & 425 & 497246 & 493 & 128554 \\
              &  [100\%] & [100\%] & [100\%] & [100\%] \\
    \hline
    \multirow{2}*{Cut 1} & 339 & 213995 & 392 & 53224 \\
    &  [79.76\%] & [43.04\%] & [79.51\%] & [41.40\%] \\
    \hline
    \multirow{2}*{Cut 2} & 189 & 10758 & 219 & 2702 \\
    &  [44.47\%] & [2.16\%] & [44.42\%] & [2.10\%] \\
    \hline
    \multirow{2}*{Cut 3} & 173 & 9218 & 200 & 2324 \\ 
    &  [40.70\%] & [1.85\%] & [40.57\%] & [1.81\%] \\
    \hline
    Significance, $\mathcal{Z}$ & \multicolumn{2}{c|}{1.802} & \multicolumn{2}{c|}{4.149} \\ \hline
    \end{tabular}}
    \caption{Cutflow table and signal significance for signal and background process at the ILC ($\sqrt{s} = 1$ TeV, $\mathfrak{L}_{\rm int}=8$ ab$^{-1}$), with unpolarized and polarized $\left(+20\%, +80\%\right)$ beam configurations. The DM signal corresponds to the benchmark: $m_{\phi} = 5$ GeV and $\Lambda = 2.5$ TeV.}
    \label{tab:cutflow}
\end{table}
\subsection{Signal Significance}
\label{sec:sign}
The statistical significance of the signal is defined as:
\begin{equation}
    \mathcal{Z} = \frac{S}{\sqrt{B}}\,,
\end{equation}
where $S$ and $B$ are the signal and background event counts, respectively. This definition quantifies how many standard deviations the signal stands out above the Poisson uncertainty in the background, $\sqrt{B}$. The signal significance after applying the final selection cut is presented in Tab.~\ref{tab:cutflow}, for the chosen benchmark point. We observe that the polarized beam setup leads to a two-fold enhancement in the signal significance compared to the unpolarized case.
\begin{figure}[htb!]
    \centering
    \includegraphics[width=0.85\linewidth]{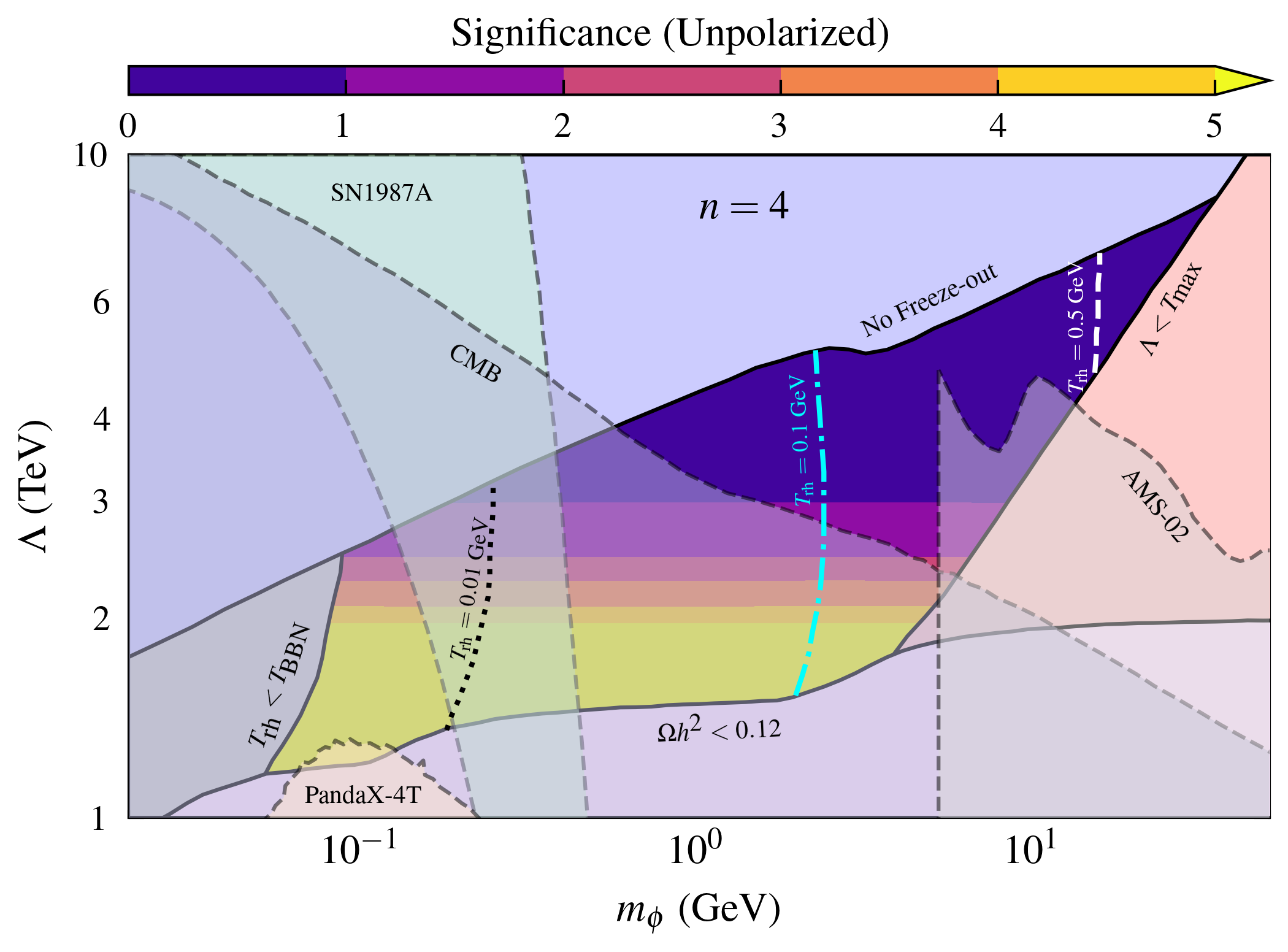}
    \includegraphics[width=0.85\linewidth]{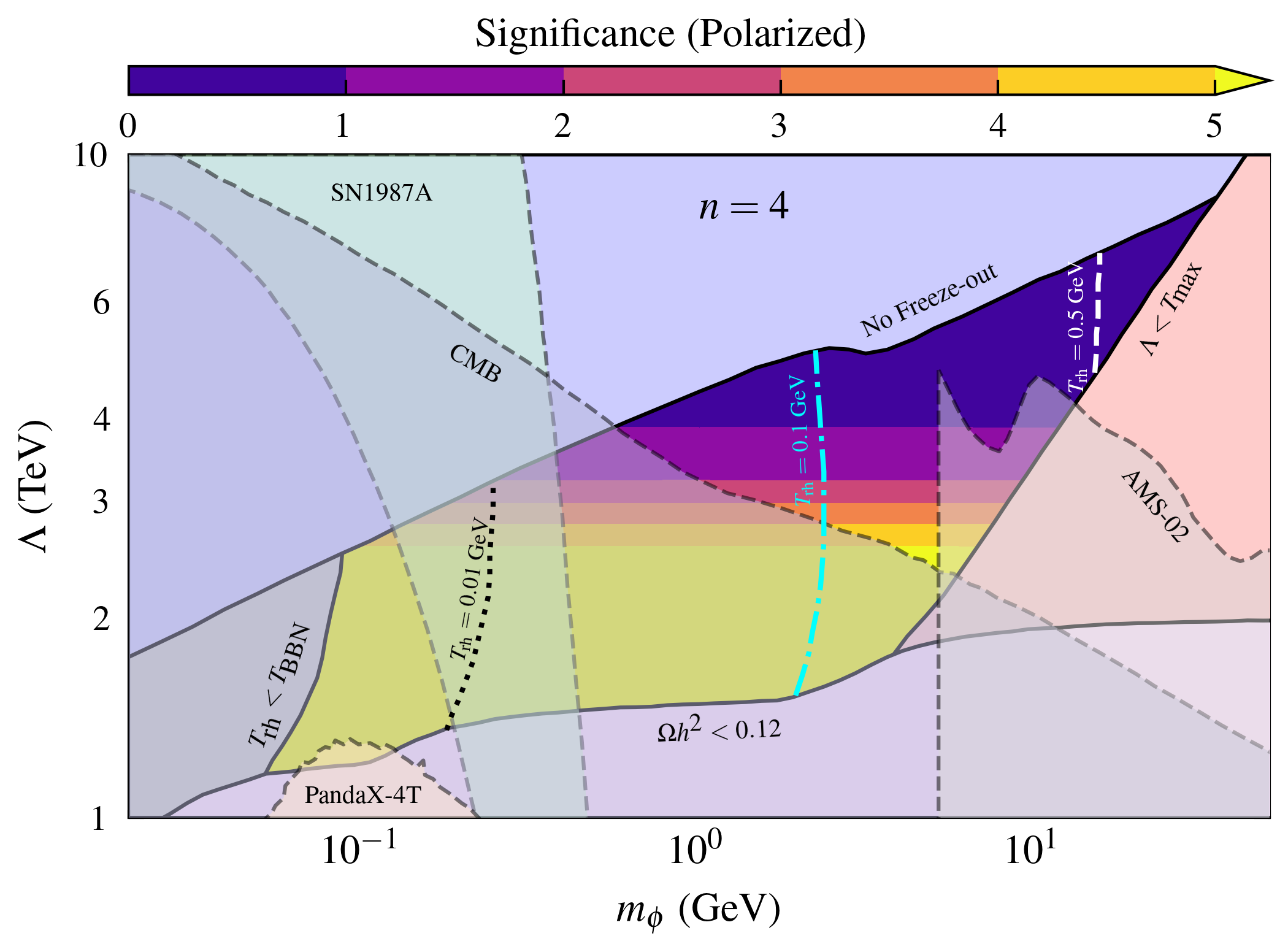}
    \caption{Signal significance on $m_{\phi}-\Lambda$ plane, for mono-$h$ signal at the ILC ($\sqrt{s} = 1$ TeV, $\mathfrak{L}_{\rm int}=8$ ab$^{-1}$). The top and bottom figures correspond to unpolarized and polarized $\left(+20\%, +80\%\right)$ cases respectively.}
    \label{fig:sig}
\end{figure}
Fig.~\ref{fig:sig} shows the signal significance in the $m_{\phi}$–$\Lambda$ plane. The various monochromatic shaded regions represent different theoretical and observational constraints within the framework. The wedge-shaped colored region corresponds to the allowed parameter space, with the color bar indicating the signal significance. The discontinuous vertical lines denote regions where the relic abundance is satisfied for specific reheating temperatures. We observe that higher reheating temperatures (e.g., $\gtrsim 500$ MeV) yield lower collider signal significance. In contrast, regions of parameter space consistent with the observed DM relic density at lower reheating temperatures can yield enhanced signal significance while remaining compatible with BBN constraints.

Thus, by performing a dedicated optimization tailored to the dimension-6 leptophilic operator, the projected data from the lepton collider can impose meaningful constraints on the DM mass versus new physics scale ($\Lambda$) parameter space. When these collider constraints are combined with bounds from indirect detection experiments such as AMS-02 and CMB, along with theoretical requirements, such as the reheating temperature satisfying $T_{\rm RH} > T_{\rm BBN}$, correct relic abundance, and the validity of the EFT framework requiring $\Lambda > T_{\rm max}$, a complete picture emerges. As illustrated in Fig.~\ref{fig:sig}, the interplay of these constraints enables us to infer viable ranges of the reheating temperature, offering novel insights into the thermal history of the early Universe in scenarios where DM freezes out during the reheating phase. We also find that $T_{\rm RH}^{}\lesssim25$ MeV are excluded (due to the DM ID bound from CMB observations) for the leptophilic operator considered in our study.

\section{Summary and Conclusion}
\label{sec:conc}
To summarize, in this analysis, we have explored the possibility that dark matter (DM) undergoes freeze-out during the reheating epoch of the early universe. Such a scenario is particularly intriguing, as reheating remains a poorly understood phase and its thermal history can crucially shape the DM relic abundance. We employ a dimension-6 leptophilic effective operator, supressed by an effective scale, $\Lambda$, to model the DM ($\phi$, scalar in nature) interactions with the visible sector, which not only makes collider detection viable through clean final states but also naturally links to early-universe processes involving leptons. We demonstrate that this framework not only modifies and often enlarges the parameter space compatible with the observed relic abundance, but also inherently carries imprints of the reheating dynamics, particularly the reheating temperature. Thus, collider detection of such a DM candidate offers a unique opportunity to gain insight into the reheating phase of the universe.

In Sec.~\ref{sec:dmprod}, we initiate our analysis by considering a simplified setup where the inflaton field, $\Phi$, decays through oscillations at the minimum of a monomial potential, thereby generating the Standard Model (SM) thermal bath and reheating the universe. The degree of the monomial potential, denoted by $n$, controls the nature and intensity of the oscillations. We examine two exclusive scenarios: fermionic reheating, where $\Phi$ decays into a pair of vector-like fermions, and bosonic reheating, where $\Phi$ decays into a pair of Higgs. In the fermionic scenario, for all values of $n$, the maximum temperature attained by the universe tends to be high. This, in order to validate the EFT framework, pushes the effective cutoff scale $\Lambda$ to very large values, often beyond the reach of current and future collider experiments. In contrast, the bosonic reheating scenario allows a viable parameter space only for $n>2$, and within a limited window of low reheating temperatures.

When incorporating the dynamics of freeze-out during reheating, we further find that scenarios with $n \geq 6$ are unable to accommodate a successful freeze-out of DM, narrowing our focus to the $n=4$ case. To delineate the collider-accessible parameter space, we impose constraints on the $m_{\phi}-\Lambda$ plane arising from the absence of freeze-out, the lower bound on reheating temperature from BBN, the validity of the effective theory, and the DM relic density under-abundance.

In Sec.~\ref{sec:constraints}, we outline the constraints on the DM parameter space arising from existing experiments and astrophysical observations. These include bounds from direct detection experiments targeting DM-electron scattering, particularly relevant for low mass DM due to the leptophilic nature of the operator under consideration, as well as indirect detection limits from positron excess measurements. Additionally, we incorporate constraints from supernova cooling and DM free streaming within the core of SN1987A, along with exclusion limits from mono-$\gamma$ searches at the LEP collider. The subset of these constraints relevant to our model and parameter choices are applied to significantly narrow down the viable region to be probed in the collider analysis.

In Sec.~\ref{sec:collider}, We demonstrate that future high-energy lepton colliders, such as the ILC, offer a promising avenue to probe DM scenarios involving leptophilic effective operators using the mono-$h$ ($h \to b\bar{b}$) channel. Compared to conventional mono-$\gamma$ or mono-$Z$ searches, the mono-Higgs channel provides improved signal-to-background separation, owing to the non-overlapping distribution of kinematic variables such as missing transverse momentum and missing energy with the dominant SM backgrounds. The contact nature of the effective operator, along with the clean collider environment and beam polarization capabilities, enhances sensitivity to otherwise elusive dark sector signatures. Our cut-based analysis at $\sqrt{s} = 1$~TeV with $\mathfrak{L}_{\rm int} = 8$~ab$^{-1}$ shows that a portion of the parameter space, corresponding to a MeV-scale reheating temperatures, can be probed at significance $> 3\sigma$. This not only provides a complementary test of freeze-out during reheating but also highlights the potential of collider signals to indirectly infer the reheating temperature.

In conclusion, by exploring the mono-$h$ channel at future lepton colliders like the ILC, we demonstrate that collider signatures of DM can act as an indirect window into the thermal history of the early universe. A statistically significant excess in this channel, within the viable parameter space shaped by freeze-out during the reheating epoch, would favor a MeV-scale reheating temperature consistent with BBN bounds. This not only expands the scope of DM searches at colliders but also provides a novel avenue to probe the dynamics of the reheating epoch, an otherwise inaccessible phase of cosmic evolution at laboratory-based experiments. The study is generic in nature and can be extended to a large class of DM models, given that the associated signatures can be probed at current and future experiments. Our analysis thus highlights the potential of collider experiments to shed light on both particle and cosmological frontiers, reinforcing the deep interplay between the early universe phenomena and collider observables.

\acknowledgments
NM and AS thank Dipankar Pradhan for his guidance with the numerical implementations and for his valuable suggestions during the preparation of this draft. NM acknowledges the Council for Scientific \& Industrial Research (CSIR), Government of India, for granting the Senior Research Fellowship. We also extend sincere thanks to Nayan Das for the many insightful discussions regarding supernova constraints on low-mass leptophilic DM.

\section*{Data Availability Statement}
The simulated datasets for this study are produced using standard high-energy physics tools, with the complete simulation procedures described in the article.

\appendix
\section{DM relic density evolution for $\boldsymbol{n=6}$ case}
\label{app:dmn6}
 \begin{figure}[h]
    \centering
    \includegraphics[width=0.60\linewidth]{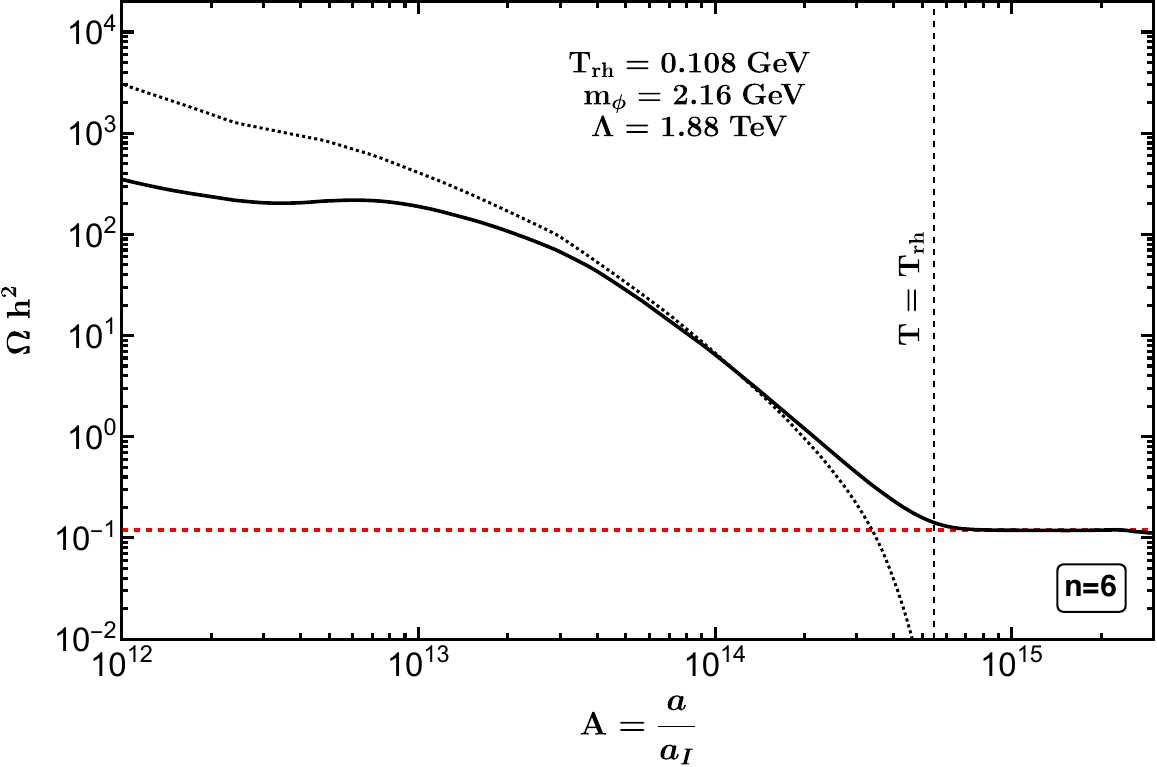}
    \caption{Same as Fig.~\ref{fig:relicDM} but for $n=6$ case.}
    \label{fig:relicDMx}
\end{figure}
Fig.~\ref{fig:relicDMx} demonstrates the DM relic evolution for the $n=6$ case, for the same DM mass and almost identical reheating temperature as in Fig.~\ref{fig:relicDM}. The correct DM relic is obtained with a higher value of $\Lambda$. This occurs because as $n$ increases, the DM abundance decreases (see Eq.~\ref{eq:Omegarh}), to offset this reduction, the DM annihilation cross-section must be lowered, which can be accomplished by increasing $\Lambda$. Also note that for $n=6$ the DM never stays in equilibrium, the DM number density crosses the equilibrium number density at a certain point in time, hence is not an example of thermal DM production similar to WIMP.  For the bosonic reheating case, the Hubble parameter in the reheating phase evolves with time as:
\begin{equation}
    \mathcal{H}(T)=\mathcal{H}(T_{\rm rh}^{})\left(\frac{T}{T_{\rm rh}^{}}\right)^{2n}\,.
\end{equation}
The above expression reveals that, for a fixed temperature, with increasing $\rm n$, the Hubble parameter increases, whereas the temperature-independent part ($s-$wave approximation) of the thermal average annihilation cross-section of the DM remains unaffected. This power-law enhancement of the Hubble parameter prevents the DM from reaching thermal equilibrium in the case of $n=6$ and higher.
\section{Feynman Diagrams for the SM Background}
\label{sec:SMBG}
\begin{figure}[htb!]
	\centering
	\includegraphics[width=0.85\linewidth]{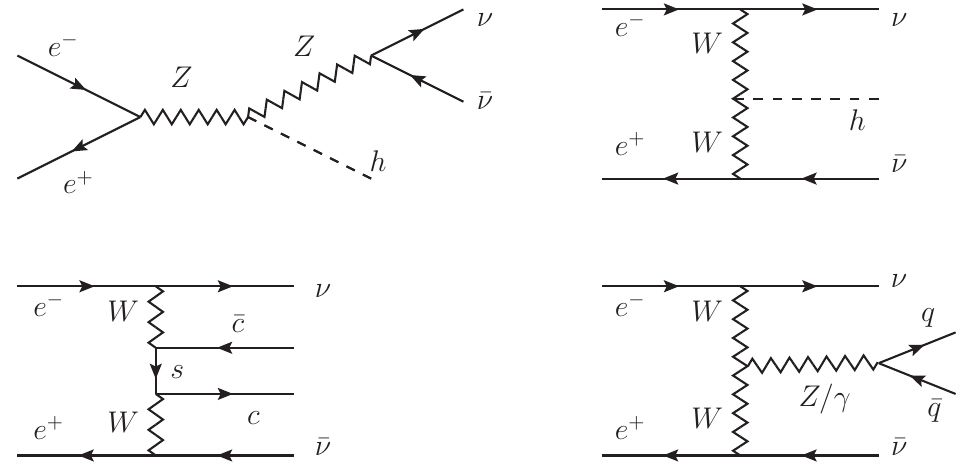}
	\caption{Representative Feynman diagrams illustrating the Standard Model background contributions to the $\nu \overline{\nu}h$ (top row) and $\nu \overline{\nu} + \text{jets}$ (bottom row) channels. Here, $q$ denotes SM quarks.}
	\label{fig:BG}
\end{figure}
\bibliographystyle{JHEP}
\bibliography{biblio.bib}
\end{document}